\documentclass[preprint,12pt,number,3p,compress]{elsarticle} 
\usepackage{framed,multirow}
\usepackage{amssymb} 
\usepackage{latexsym}
\usepackage{graphicx}
\usepackage{caption}
\usepackage{dcolumn}
\usepackage{bm}
\usepackage{float}
\usepackage{color}
\usepackage{xcolor}
\usepackage{tcolorbox}
\usepackage{amsmath} 
\usepackage[a]{esvect} 
\usepackage{algorithm}
\usepackage[noend]{algpseudocode}
\usepackage{amsfonts}
\usepackage{placeins}
\usepackage{verbatim} 
\usepackage{chemformula} 

\everymath{\displaystyle}

\definecolor{red}{RGB}{255,0,0}

\journal{International Journal of Hydrogen Energy}

\begin{document}
\captionsetup[figure]{labelfont={bf},labelformat={default},labelsep=period,name={Fig.}}

\begin{frontmatter}


\title{Grand Canonical Monte Carlo Simulations of Hydrogen and Methane 
Storage Capacities of Two Novel Al-nia MOFs at Room Temperature}

\author[Valladolid]{A. Granja-DelRío}
\author[Valladolid]{I. Cabria\corref{cor1}}
\ead{ivan.cabria@uva.es}
\cortext[cor1]{Corresponding author. 
Tel.: +34 983 423141; Fax: +34 983 423013}
\address[Valladolid]{Departamento de F\'isica Te\'orica, At\'omica y 
\'Optica, Universidad de Valladolid, 47011 Valladolid, Spain}


\begin{abstract}

Novel materials capable of storing hydrogen or/and methane at high 
gravimetric and volumetric densities are required for hydrogen vehicles 
to be widely employed as a clean alternative to fossil-based vehicles. 
Metal-Organic Frameworks (MOFs) are considered as promising candidates 
to achieve the Department Of Energy (DOE) targets for both, hydrogen 
and methane storage. Using Grand Canonical Monte Carlo (GCMC) 
simulations, the hydrogen and methane gravimetric and volumetric 
storage capacities of two recently synthesized Al-nia MOFs have been 
studied. Their storage capacities have been compared with the storage 
capacities of other Al-based MOFs and classical and well-known MOFs, 
such as IRMOF-5. The two novel Al-nia MOFs have shown high hydrogen and 
methane gravimetric and volumetric storage capacities at room 
temperature and moderate pressures, 25-35 MPa, comparable or higher 
than the storage capacities of classical and Al-based MOFs.

\end{abstract}

\begin{keyword}
Hydrogen Storage, Hydrogen Physisorption, Methane Storage, Al-based 
MOFs, Grand Canonical Monte Carlo Simulations
\end{keyword}

\end{frontmatter}

\section{Introduction}

Humankind evolution has led to important environmental threats and a major increase of energy consumption. Fuel combustion produces several harmful pollutants such as carbon dioxide or carbon monoxide. Adopting new strategies to replace fossil fuels could solve this problem. Hydrogen is considered a very promising candidate to solve these issues. Basically, hydrogen reacts with atmospheric oxygen producing an electric current and emitting water. However, on-board hydrogen storage at room temperature and moderate pressures remains the principal obstacle to implement 
large fleets of fuel cell hydrogen vehicles \cite{breeze18}. Nowadays, hydrogen storage in vehicles is achieved by compressing 
gas at very high pressures. Decreasing these pressures will reduce the costs of these vehicles and refueling stations. The Department of Energy established two main targets for an on-board hydrogen storage system, 
to achieve before 2025: (a) a gravimetric capacity of at least 5.5 wt. \% and (b) a volumetric capacity of at least 0.040 kg of hydrogen per liter \cite{ahg+18,doevehicle20202025andultimate}. The goal of these 
targets is to obtain a hydrogen vehicle equivalent to the present 
average fossil fuel-based vehicle.

On the other hand, while the performance of hydrogen-powered vehicles 
is being improved, it might be necessary a transitional bridge between 
gasoline and hydrogen-based automobiles. Natural gas feels a good 
alternative since it is widely distributed, has huge reserves, low cost 
and is cleaner than oil. The reason is that the main component of 
natural gas, methane, has the highest ratio of hydrogen to carbon of 
all fossil fuels. Compared to coal, natural gas can reduce \ch{CO2} 
emissions by 50 \% \cite{ch4maurin}. Methane has higher gravimetric 
capacity than gasoline but, the challenge is to increase the volumetric 
density at ambient conditions. The main problem is, again, the storage. 
Natural gas is stored onboard the vehicle in compressed or liquefied 
form. These need expensive and bulky storage. The Advanced Research 
Projects Agency - Energy (ARPA-E) of the DOE set the targets for 
methane storage to lead the research on adsorbent materials for methane 
storage. The volumetric capacity needs to achieve 0.250 kg of methane/L 
and the gravimetric capacity 0.5 g/g or 33.33 wt. \% at room 
temperature and moderate pressures \cite{movetargets12}.

Three main different types of hydrogen and methane storage have been 
proposed: liquid, compressed and solid. The last one includes the storage 
by physisorption on solid porous materials. That storage method allows 
storing more gas at low and moderate pressures ($\leq$ 25-35 MPa) than the 
compressed method \cite{cngtanks,h2tanks}. Development of efficient 
adsorbent solid porous materials is the key to achieve an efficient on-board 
storage technology. Many different porous materials have been evaluated as 
promising gas storage materials. Activated carbons form a wide group of 
materials with a well-developed porosity, light weight and large hydrogen 
and methane storage capacities \cite{zzc23,am23,bm22,ajpco20}. Covalent 
organic frameworks (COFs) represent another class of porous materials that 
are constructed through the covalent bonding of organic building blocks. 
COFs exhibit permanent porosity and possess tunable structures, making them 
highly appealing for gas storage applications \cite{dbvrv23,dzwty23,wll23}. 
Polymers of Intrinsic Microporosity (PIMs) possess a unique molecular 
structure that inherently incorporates microporosity, resulting in porous 
materials with remarkably high internal surface areas. This inherent 
characteristic grants PIMs the ability to display enhanced capacities for 
gas sorption \cite{mac22,ckiwf22,trf+20}.

In the last decades, many studies on adsorbed hydrogen and natural gas have 
focused on Metal-Organic Frameworks (MOFs), which are a group of promising 
solid porous materials for hydrogen and methane storage \cite{ltyw23,lst+22,
Li2022,spst21,yjllj20,csmo20,ch4chen14}. They are made of metal ions or 
metal containing clusters (secondary building blocks) linked by organic 
ligands to form 2D or 3D structures. The huge amount of organic linkers 
and the large variety of metal ions and metal clusters conduct to many 
different MOFs with tunable properties.

Recently, Alezi et al. \cite{alniamof2022} have synthesized two novel 
Al-nia MOFs which showed high porosity. Aluminum is an abundant element 
and hence, the synthesis of these MOFs is not expensive. Aluminum is 
also a light element and this implies, in general, high gravimetric 
storage capacities. The aims of this investigation are to predict and 
to study the hydrogen and methane storage capacities of these promising 
novel MOFs. Grand Canonical Monte Carlo (GCMC) simulations of the 
hydrogen and methane storage capacities of those two recently 
synthesized Al-nia-MOFs and of other MOFs have been performed. The 
storage capacities of these new materials have been compared with the 
capacities of other Al-based MOFs and of a set of classical and 
well-known MOFs. The storage capacities of all the MOFs simulated have 
been also analyzed as functions of the porosity and density of the 
MOFs.


\section{Methodology}

The GCMC simulations of the hydrogen and methane gravimetric and 
volumetric storage capacities of the two novel Al-nia-MOFs 
(Al-nia-MOF-1 and Al-nia-MOF-2), of a set of 15 Al-based 
MOFs, selected from the Cambridge Crystallographic Database Centre 
(CCDC) \cite{ccdc} and of a set of 21 
classical MOFs have been performed at 298.15 K and pressures between 0.5 
and 35 MPa. In addition, for Al-nia-MOF-1, \ch{CH4} storage 
simulations at 298 K and 0.5 and 8 MPa were performed in order to 
compare the theoretical results with those from the experiment made 
by Alezi et al. \cite{alniamof2022}.

The number of iterations in each GCMC simulation was ten millions. The 
first five millions were set to reach the equilibrium and the rest 
were employed to calculate the capacities. The Metropolis algorithm was 
used in each iteration \cite{metropolis87}. The probabilities of the 
possible changes or moves were distributed as follows: 20 \% consisted 
on the movement of one molecule, 40 \% on the removal of one molecule 
and 40\% for the addition of one molecule. The simulations were 
performed using an in-house code.

The chemical potential used in the simulations was obtained from the 
Soave-Redlich-Kwong (SRK) equation of state \cite{soave72}. The values 
for the dimensionless acentric factor $\omega$, the critical pressure 
P$_c$ and the critical temperature T$_c$ of hydrogen and methane used 
in the SRK equation are those published by Zhou and Zhou \cite{zz01} 
and Xu et al. \cite{xdy12}, respectively.

Lennard-Jones (LJ) interaction potentials \cite{lennard24} have been 
used to simulate the interactions between the atoms of the MOFs and the 
molecules and between the molecules. The LJ coefficients used in the 
present simulations can be found in Table~\ref{coeffsLJ}.

\begin{table}[ht]
\begin{center}
\caption{\noindent
\label{coeffsLJ}
Lennard-Jones coefficients $\sigma$ and $\epsilon$ of the molecules 
and of the atoms of the MOFs studied in the present GCMC simulations.}
\bigskip
\begin{tabular}{cccccc}
\hline
Atom or molecule & $\sigma$(\AA{}) & $\epsilon$(eV) & Source\\
\hline
Al & 2.574 & 0.507220 & Filippova et al. \cite{fkp15} \\
Br & 3.519 & 0.016043 & Mayo et al. \cite{mog90} \\
C  & 3.400 & 0.003744 & Tu et al. \cite{txw+09} \\
Cl & 3.350 & 0.014951 & Singer et al. \cite{sts77} \\
Cu & 2.297 & 0.520310 & Filippova et al. \cite{fkp15} \\
H  & 2.846 & 0.000659 & Mayo et al. \cite{mog90} \\
N  & 3.310 & 0.003214 & Cheung et al. \cite{cp75} \\
O  & 3.033 & 0.004150 & Mayo et al. \cite{mog90} \\
S  & 3.590 & 0.014916 & Mayo et al. \cite{mog90} \\
Zn & 0.998 & 0.008291 & Soper et al. \cite{soper04} \\
\ch{H2} & 2.970 & 0.002870 & Rzepka et al. \cite{rlc98} \\
\ch{CH4} & 3.730 & 0.012748 & Jorgensen et al. \cite{jms84} \\
\hline
C-\ch{H2} & 3.190 & 0.002628 & Rzepka et al. \cite{rlc98}\\
\hline
\end{tabular}
\end{center}
\end{table}

The LJ coefficients of the interaction between an atom and a molecule 
(hydrogen or methane) have been calculated using the Good-Hope 
\cite{gh70} and the Berthelot \cite{berthelot1898} combining rules for 
the $\sigma$ and $\epsilon$ coefficients of the LJ potentials, 
respectively. In the case of the C-\ch{H2} interaction, the LJ 
coefficients published by Rzepka et al. \cite{rlc98} have been used. 
The cutoff radius of the LJ interaction potential was 20 and 7.5 \AA{} 
for the interactions with \ch{H2} and methane, respectively. These 
values of the cutoff radii were obtained after carrying out several 
tests. To take into account the quantum effects, the Feynman-Hibbs 
\cite{fh65} quantum correction has been used. 

In order to establish a comparison of interaction potentials, an alternative 
set of LJ coefficients, the MDT set, (MDT: Mayo, Darkrim and TraPPE) has 
been selected. This second set uses the LJ coefficients published by Mayo et 
al. \cite{mog90} for the atoms, the LJ coefficients by Darkrim and Levesque 
\cite{dl98} for the hydrogen molecule and the LJ coefficients of TraPPE 
\cite{trappe} for methane. These coefficients can be found in 
Table~\ref{coeffsLJMDT}.

\begin{table}[ht]
\begin{center}
\caption{\noindent
\label{coeffsLJMDT}
MDT set of Lennard-Jones coefficients $\sigma$ and $\epsilon$ of the 
molecules and of the atoms of the MOFs studied in the present GCMC 
simulations.}
\bigskip
\begin{tabular}{cccccc}
\hline
Atom or molecule & $\sigma$(\AA{}) & $\epsilon$(eV) & Source \\
\hline
Al               &  3.911          &  0.013442      &  Mayo et al. \cite{mog90}\\
Br               &  3.519          &  0.016043      &  Mayo et al. \cite{mog90} \\
C                &  3.473          &  0.004124      &  Mayo et al. \cite{mog90}\\
Cl               &  3.519          &  0.012284      &  Mayo et al. \cite{mog90}\\
Cu               &  3.093          &  0.002168      &  Mayo et al. \cite{mog90}\\
H                &  2.846          &  0.000659      &  Mayo et al. \cite{mog90}\\
N                &  3.263          &  0.003356      &  Mayo et al. \cite{mog90}\\
O                &  3.033          &  0.004150      &  Mayo et al. \cite{mog90}\\
S                &  3.590          &  0.014916      &  Mayo et al. \cite{mog90}\\
Zn               &  4.045          &  0.002385      &  Mayo et al. \cite{mog90}\\
\ch{H2}          &  2.958          &  0.003163      &  Darkrim and Levesque \cite{dl98}\\
\ch{CH4}         &  3.730          &  0.012754      &  TraPPE \cite{trappe}\\
\hline
\end{tabular}
\end{center}
\end{table}

\subsection{Calculation of the Porosity}

The porosity is the ratio of the available volume and the volume of the 
simulation cell of a MOF\@. This is a dimensionless magnitude. The 
difference between the volume of the simulation cell, $V$, and the 
volume occupied by the atoms of the MOF, $V_{occ}$ is the available 
volume to a gas molecule.

To calculate the volume $V_{occ}$, a grid of $n$ points is created in 
the simulation cell. The volume $V_{occ}$ is given by the $V ni/n$, 
where $ni$ is the number of grid points that are inside the sphere of 
some atom of the simulation cell. The radius $r$ of an atom is 
$\sigma(atom, molecule)$ and the volume of the sphere of an atom is 
given by 4$\pi r^3$/3. The LJ interaction potential energy is zero at 
an atom-molecule distance equal to $\sigma(atom, molecule)$. The 
available volume, the volume occupied by the atoms and the porosity 
depend on the type of gas molecule, due to this definition of the 
radius of an atom interacting with a molecule. The experimental porosity 
also depends on the type of gas.

\subsection{Definitions of the Storage Capacities}

The total (hydrogen or methane) volumetric storage capacity, $v_c$, 
also called the density of stored (hydrogen or methane), is defined in 
these simulations by

\begin{equation}
\label{vc}
v_c = \frac{M_g}{V} \;,
\end{equation}

\noindent
where $M_g$ is the mass of gas (hydrogen or methane) stored in the 
simulation cell and $V$ is the volume of the simulation cell. 
Throughout this paper, the units of $v_c$ are kg of \ch{H2}/L or kg of 
\ch{CH4})/L.

The total (hydrogen or methane) gravimetric storage capacity, $g_c$, 
is defined by

\begin{equation}
\label{gc}
g_c = \frac{100 M_g}{M_g + M_{ads}} \;,
\end{equation}

\noindent
where $M_{ads}$ is the mass of the adsorbent solid porous material of 
the simulation cell. The units of the total gravimetric capacity are 
wt. \%.

The usable, also called delivery or working, mass of gas (hydrogen or 
methane) stored at a $P$ and $T$ is the difference between the total 
mass of gas stored at $P$ and $T$ and the total mass of gas stored at 
the depletion pressure and $T$ \cite{ahg+18,doevehicle20202025andultimate,bwf+19,sh16}. 
The usable (hydrogen or methane) volumetric and gravimetric capacities are 
given by Eqs.~\ref{vc} and~\ref{gc}, using the usable mass of gas stored. 
Only the total and usable storage capacities will be presented and analyzed 
in this research.

\subsection{Simulation Cells}

The cells of the MOFs have been obtained from their files in the 
Crystallographic Information File (CIF) format on the CCDC database 
\cite{ccdc}. The set of 21 classical MOFs is composed by IRMOF-1 to IRMOF-20 
(except 13, 17 and 19), HKUST-1, MOF-177, NU-111 and NU-125 (IRMOF: 
Isoreticular Metal-Organic Framework). 

The MOF CCDC subset contains 106 Aluminum-based MOFs. The porosity and 
the density of the 106 were calculated. Among all of them, the ones 
with porosity above 40 \% and density below 70 \% were 
selected to make GCMC simulations and to compare their GCMC storage 
capacities with those of the two novel MOFs. Usually, MOFs with those 
values of porosity and density have high storage capacities. This 
process led to a set of 15 selected Al-based MOFs whose CCDC database 
identifiers are: GUFYUE, MUZKEZ, OTECEX, OTEDOI, OTEDUO, OYEWEW, 
QOXQUS, QOXRAZ, QOXRIH, QOXSAA, QOXSUU, QOXTOP, QOXVEH, RIXPIZ and 
WOFQEQ.

\begin{figure}[ht]
\vspace{0.0cm}
\begin{center}
  \leavevmode
  \put(-220,   0){%
  \includegraphics[width=8cm,angle=0,clip]{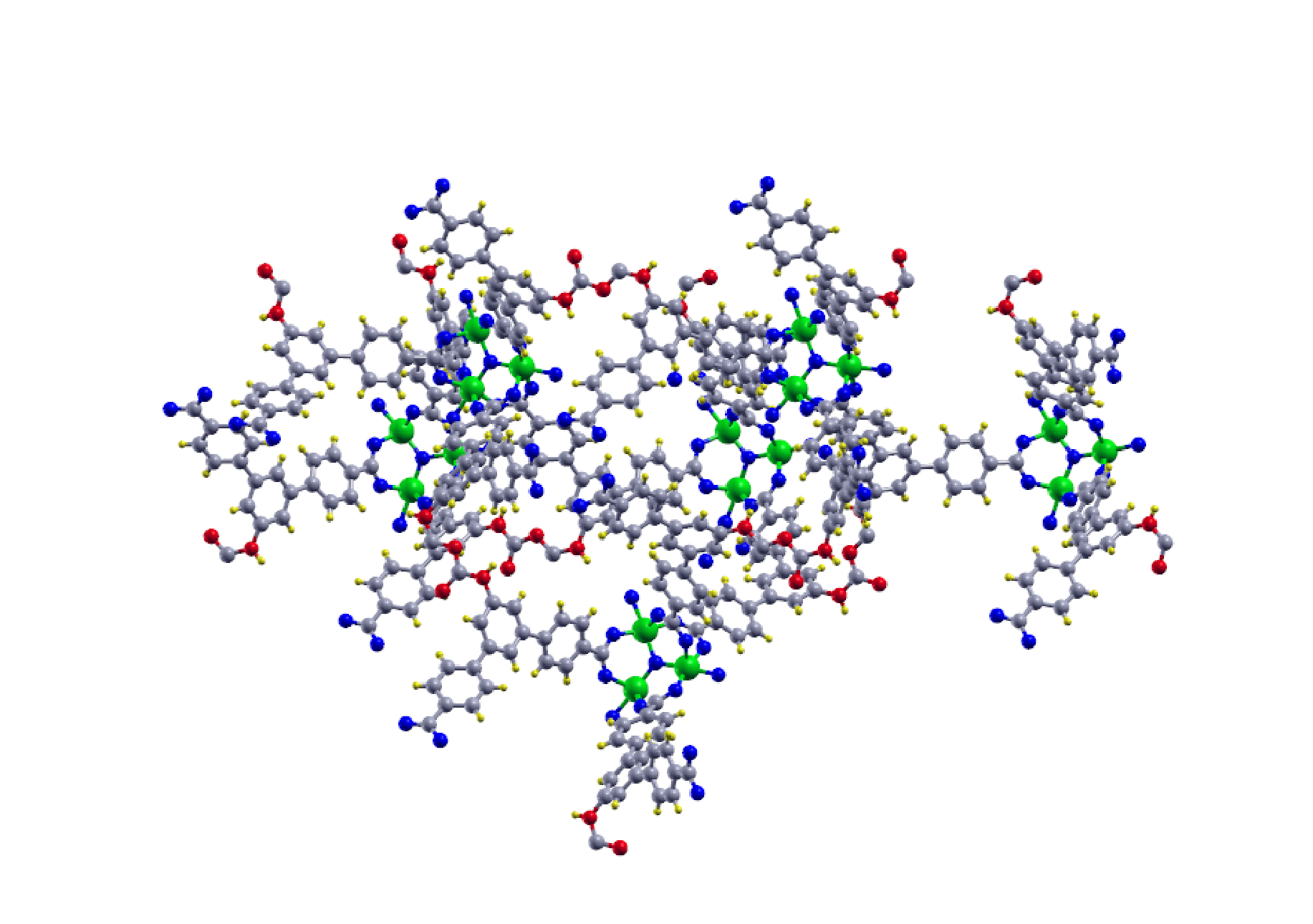}
  }
  \put(   0,   0){%
  \includegraphics[width=8cm,angle=0,clip]{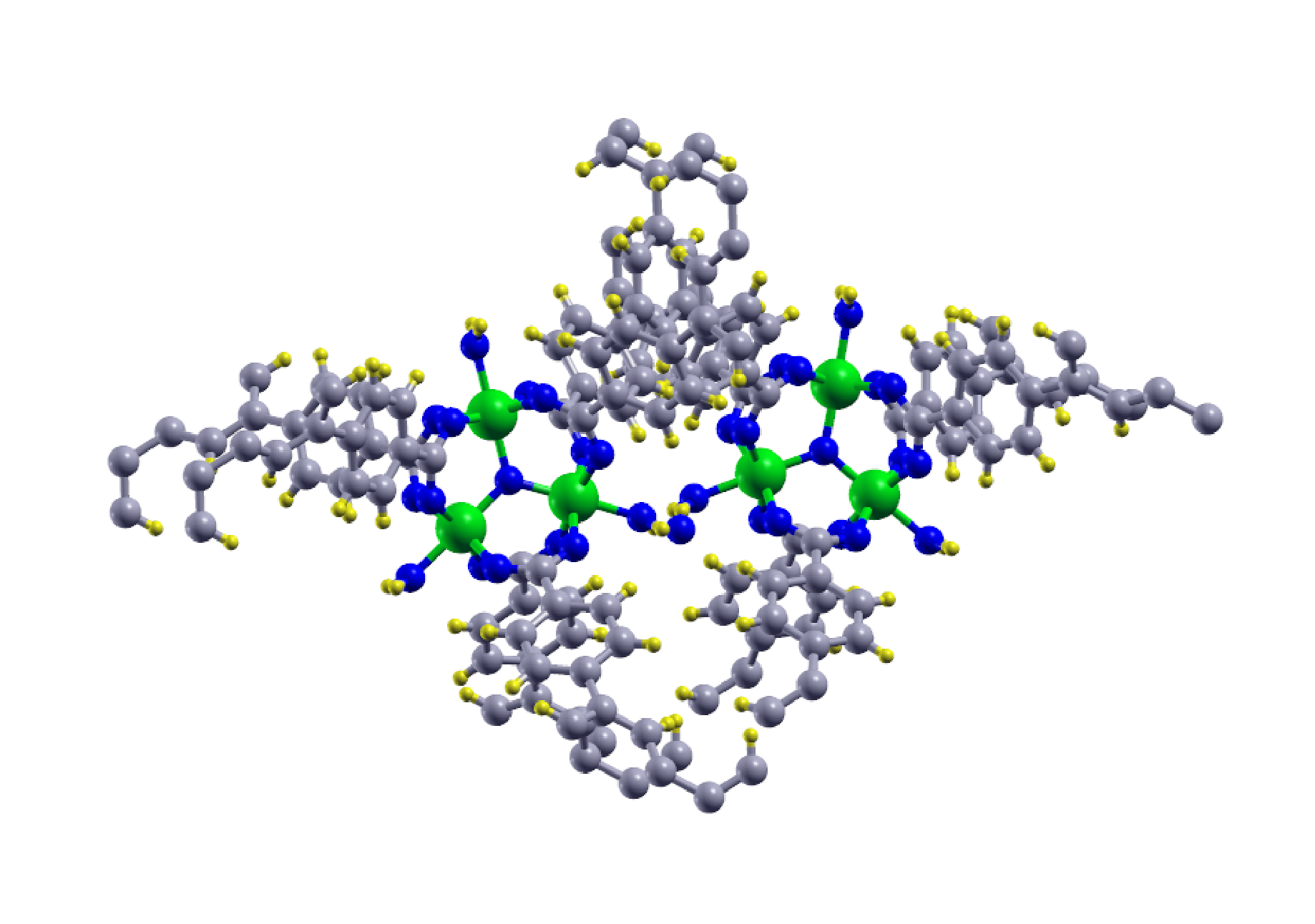}
  }
\end{center}
\vspace{-0.5cm}
\caption{Simulation cells of the two novel MOFs: Al-nia-MOF-1 and 
Al-nia-MOF-2 (left and right panels, respectively). Oxygen, carbon, 
hydrogen, aluminum and nitrogen atoms are represented by blue, gray, 
yellow, green and red balls, respectively.}
\label{cells}
\end{figure}


\section{Results}

\subsection{Hydrogen Storage Capacities}

\begin{figure}[htb]
\vspace{0.0cm}
\begin{center}
  \leavevmode
  \put(-220, 150){%
	\includegraphics[width=7cm]{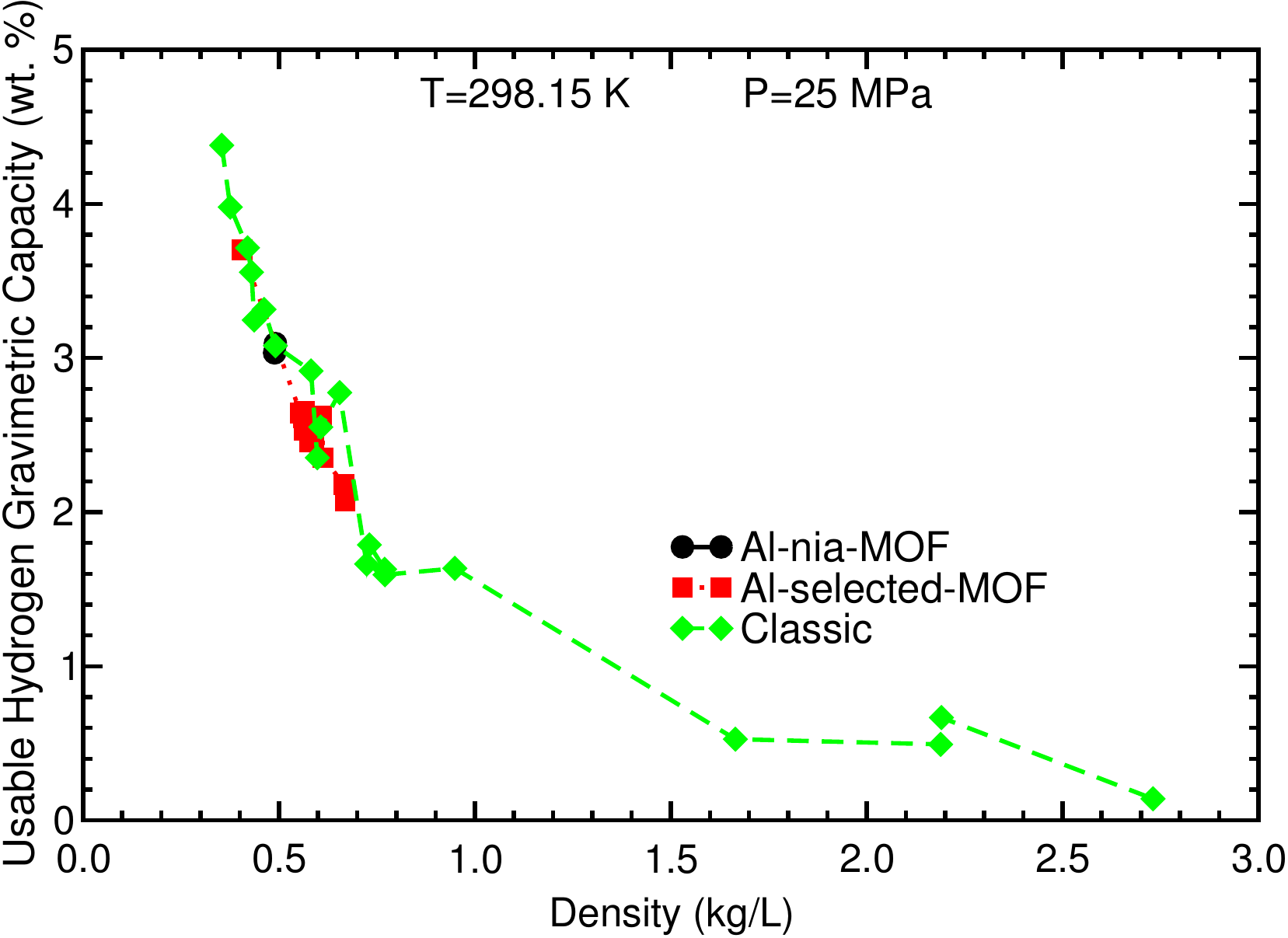}
  }
  \put(   0, 150){%
	\includegraphics[width=7cm]{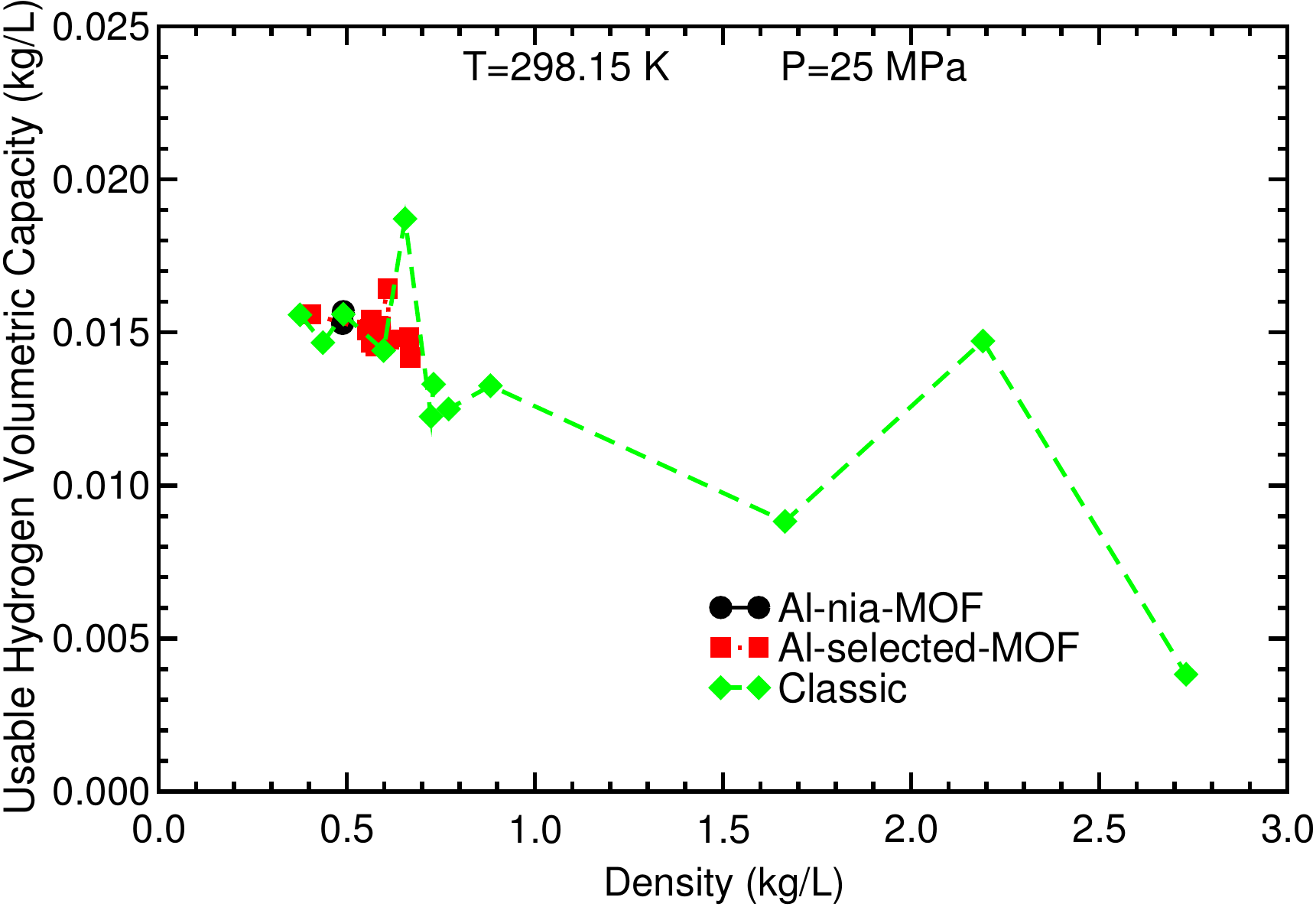}
  }
  \put(-220,   0){%
	\includegraphics[width=7cm]{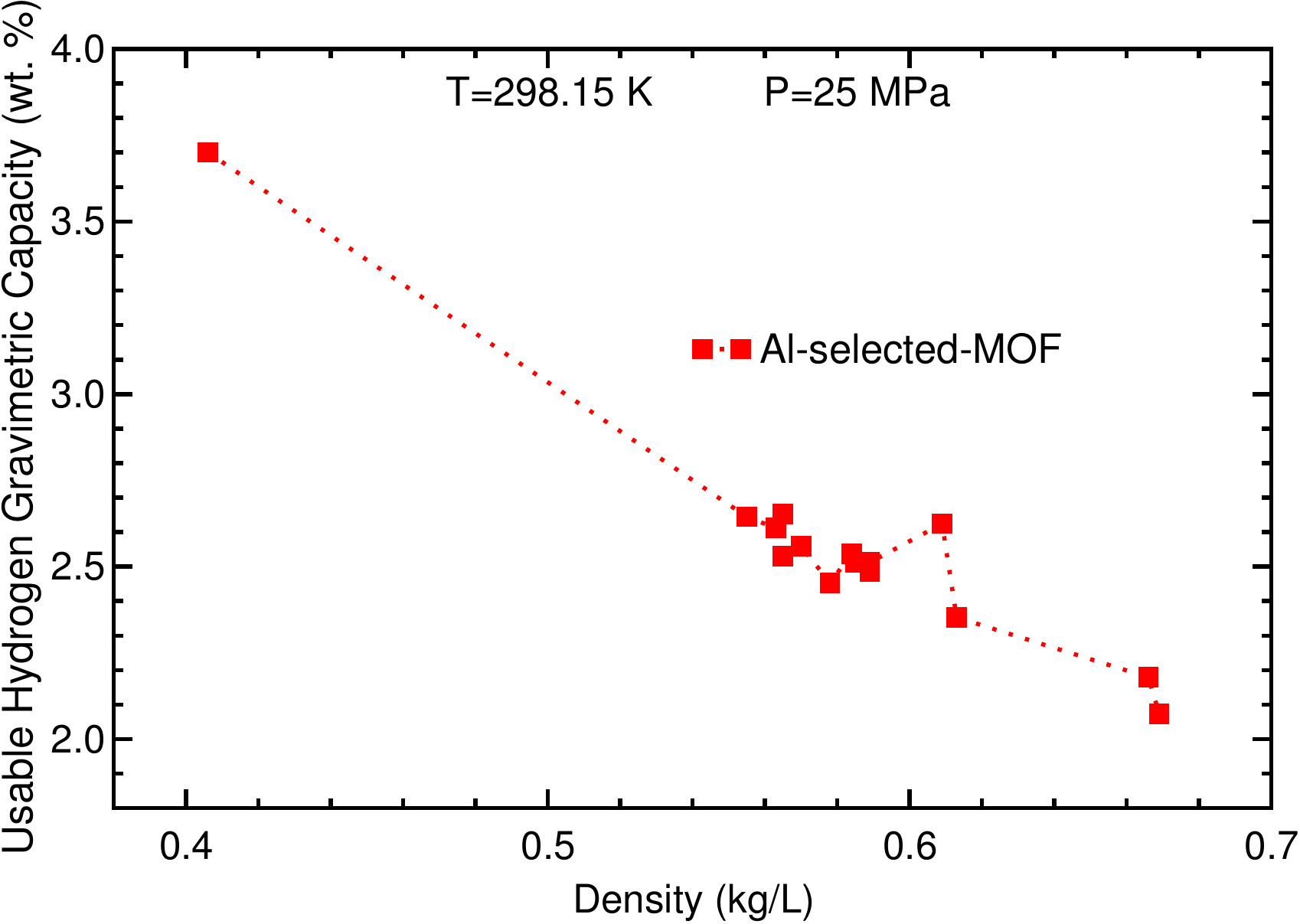}
  }
  \put(   0,   0){%
	\includegraphics[width=7cm]{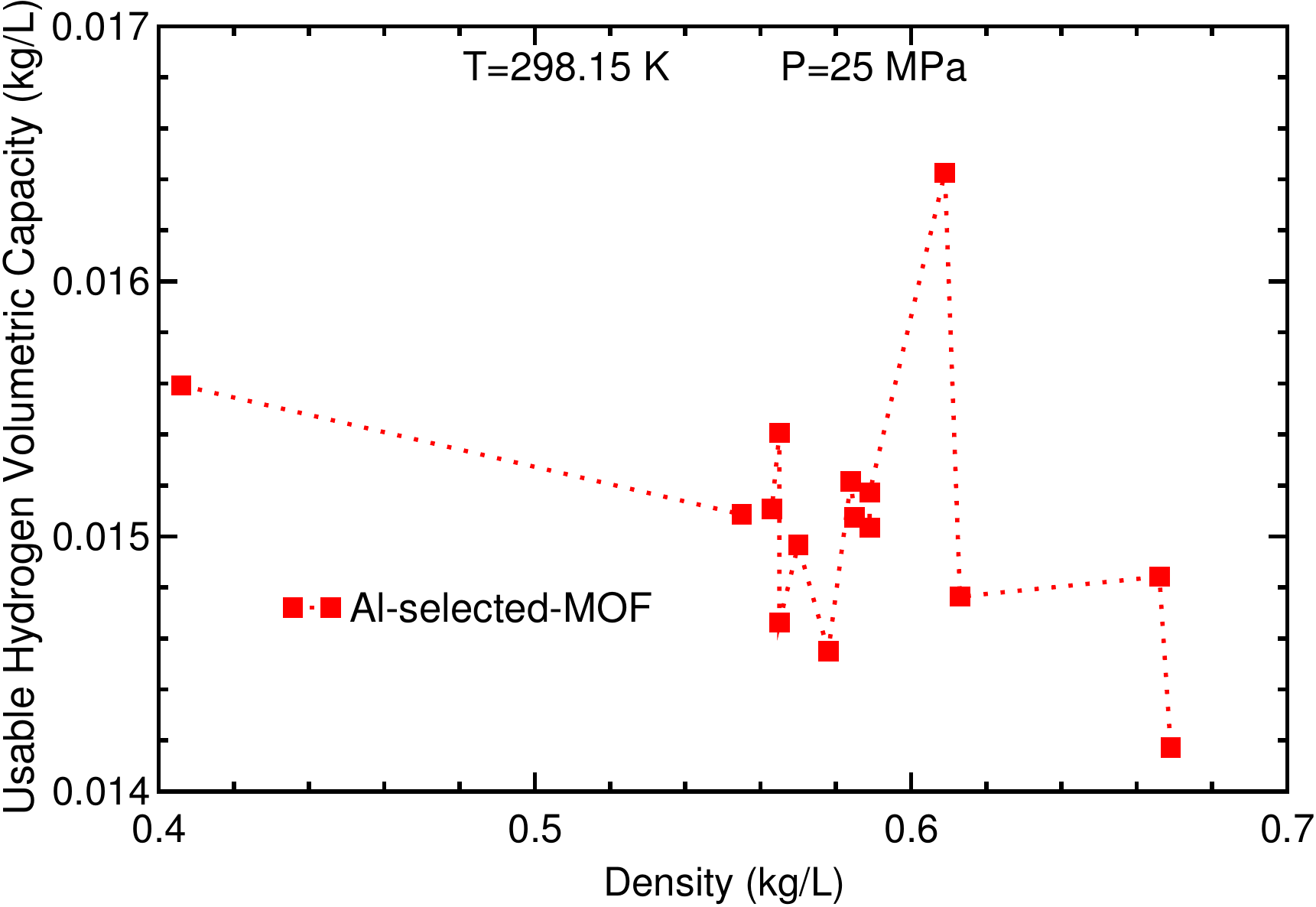}
  }
\end{center}
\vspace{-0.5cm}
\caption{Hydrogen gravimetric and volumetric usable capacity at 298.15 K and 
25 MPa vs density of Al-nia, the selected Al-based and the classical MOFs. 
Low panels show a zoomed-in view of the capacities of the selected Al-based 
MOFs.}
\label{h2capvsd}
\end{figure}

\begin{figure}[htb]
\vspace{0.0cm}
\begin{center}
  \leavevmode
  \put(-220, 150){%
	\includegraphics[width=7cm]{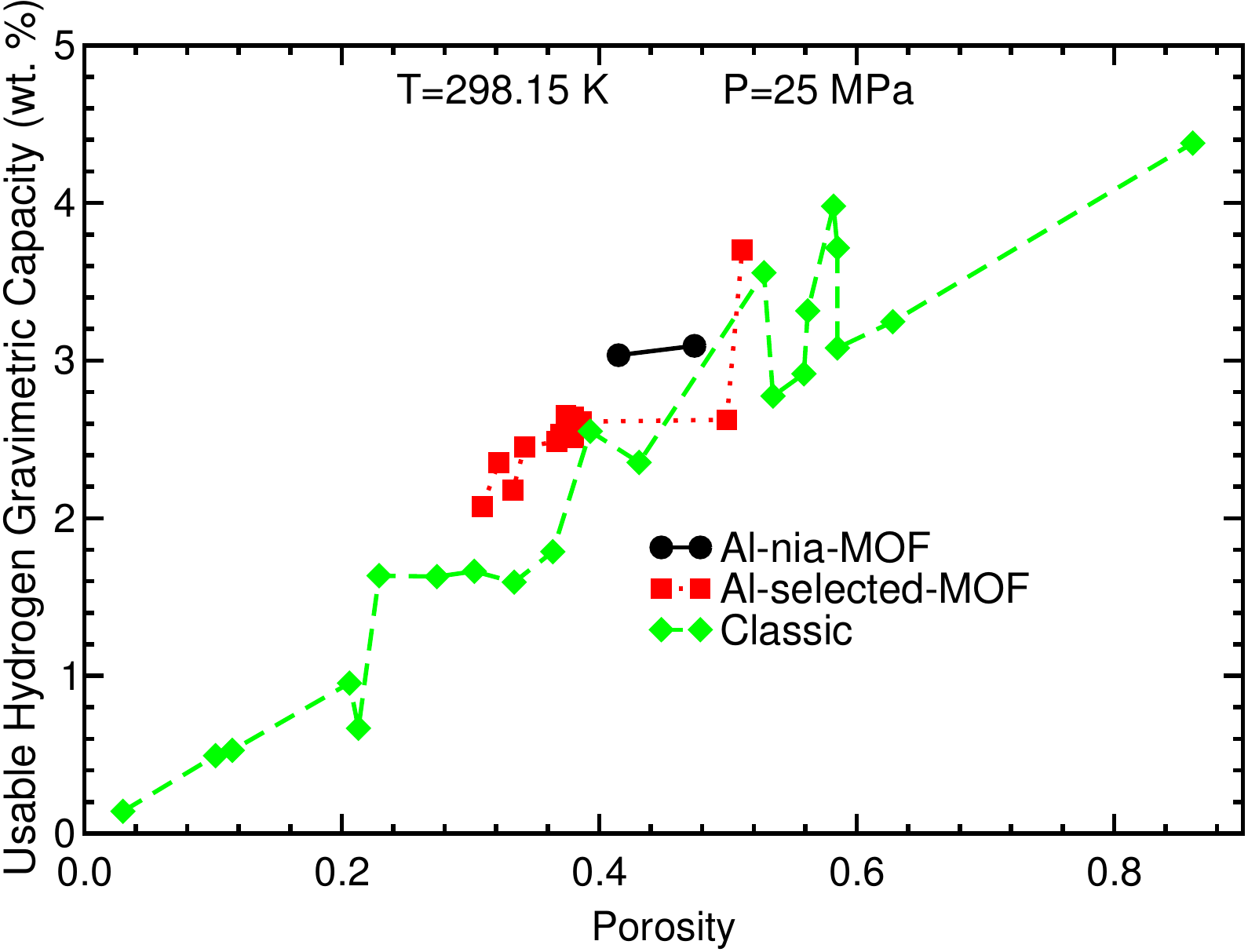}
  }
  \put(   0, 150){%
	\includegraphics[width=7cm]{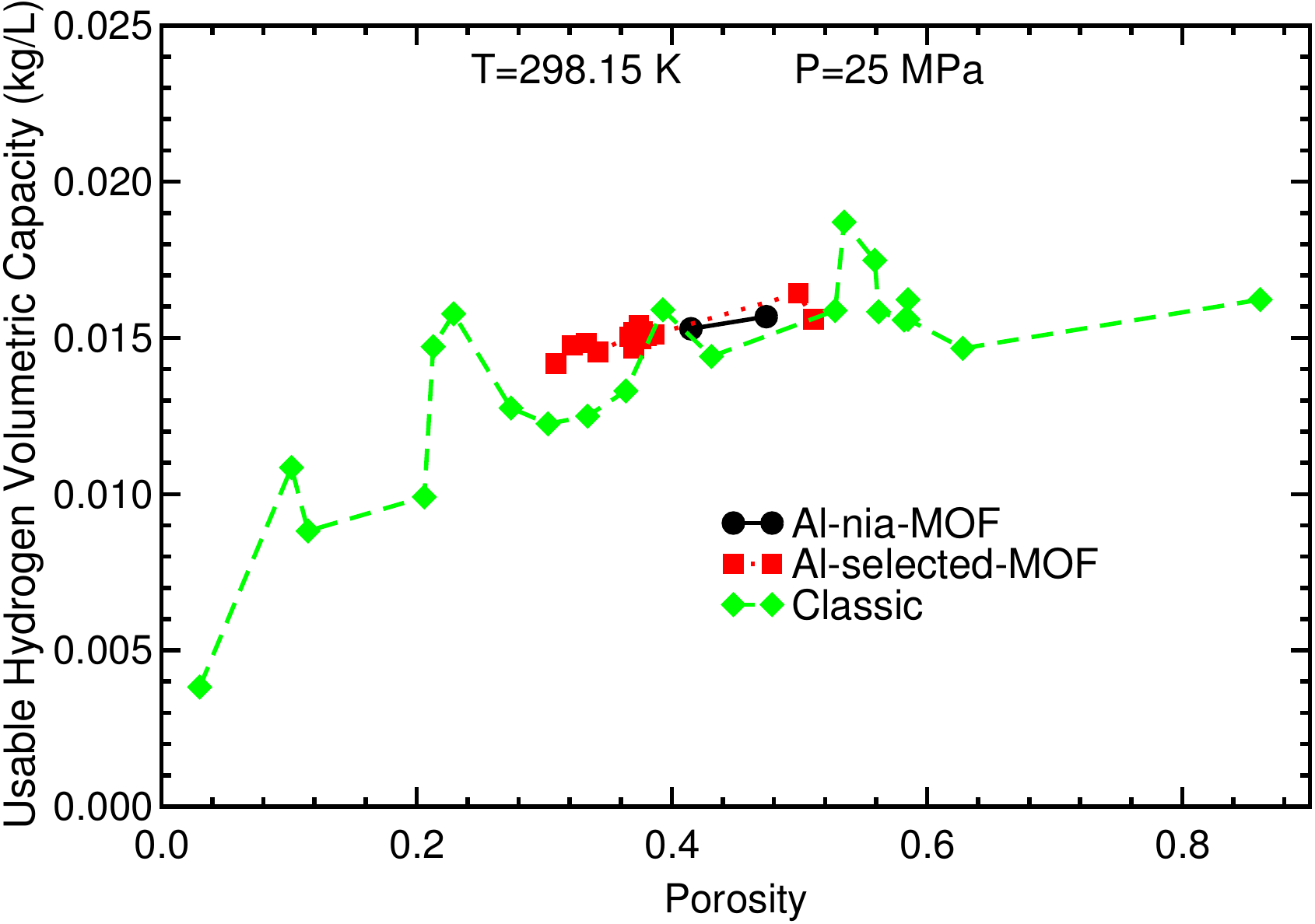}
  }
  \put(-220,   0){%
	\includegraphics[width=7cm]{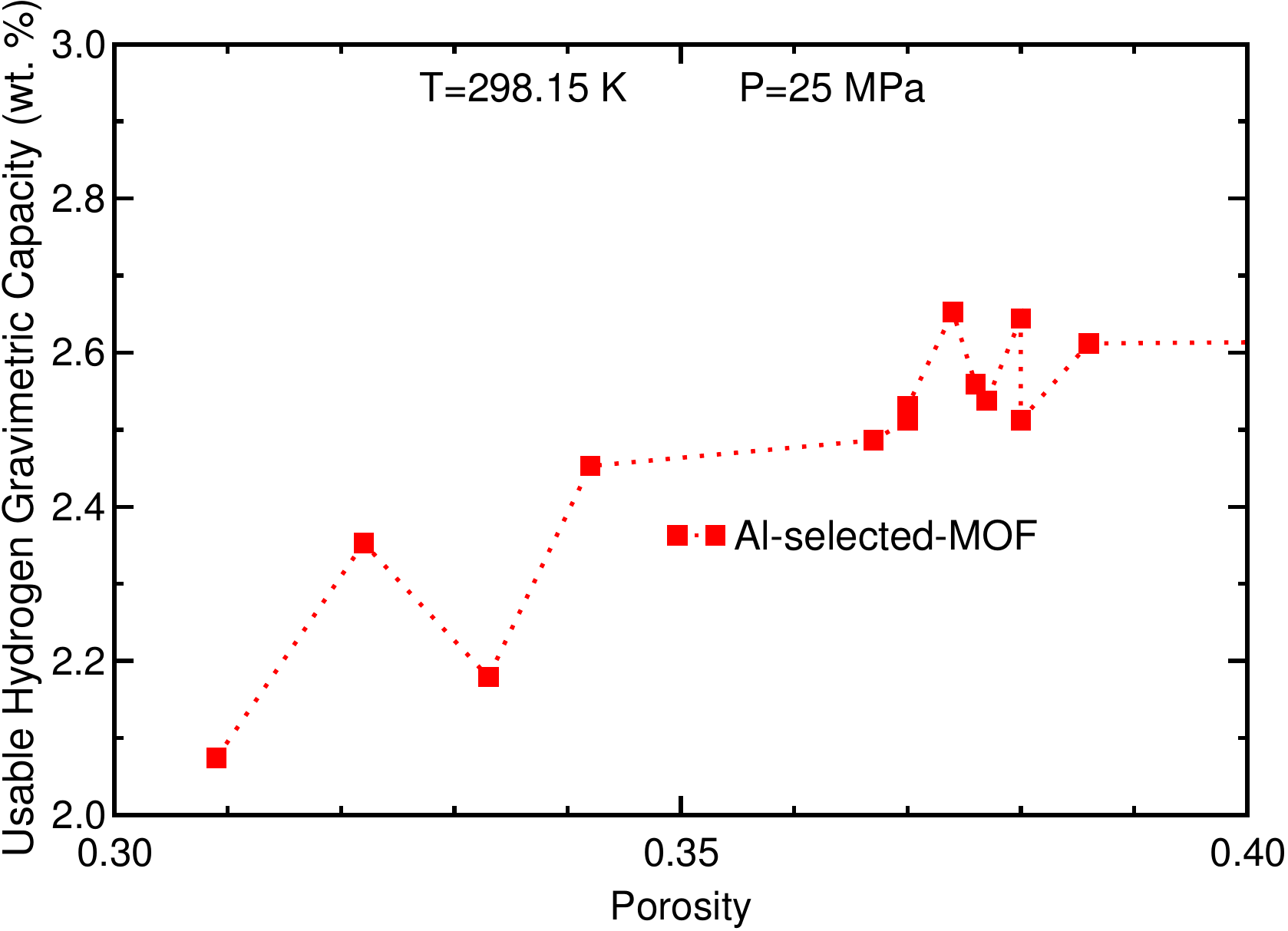}
  }
  \put(   0,   0){%
	\includegraphics[width=7cm]{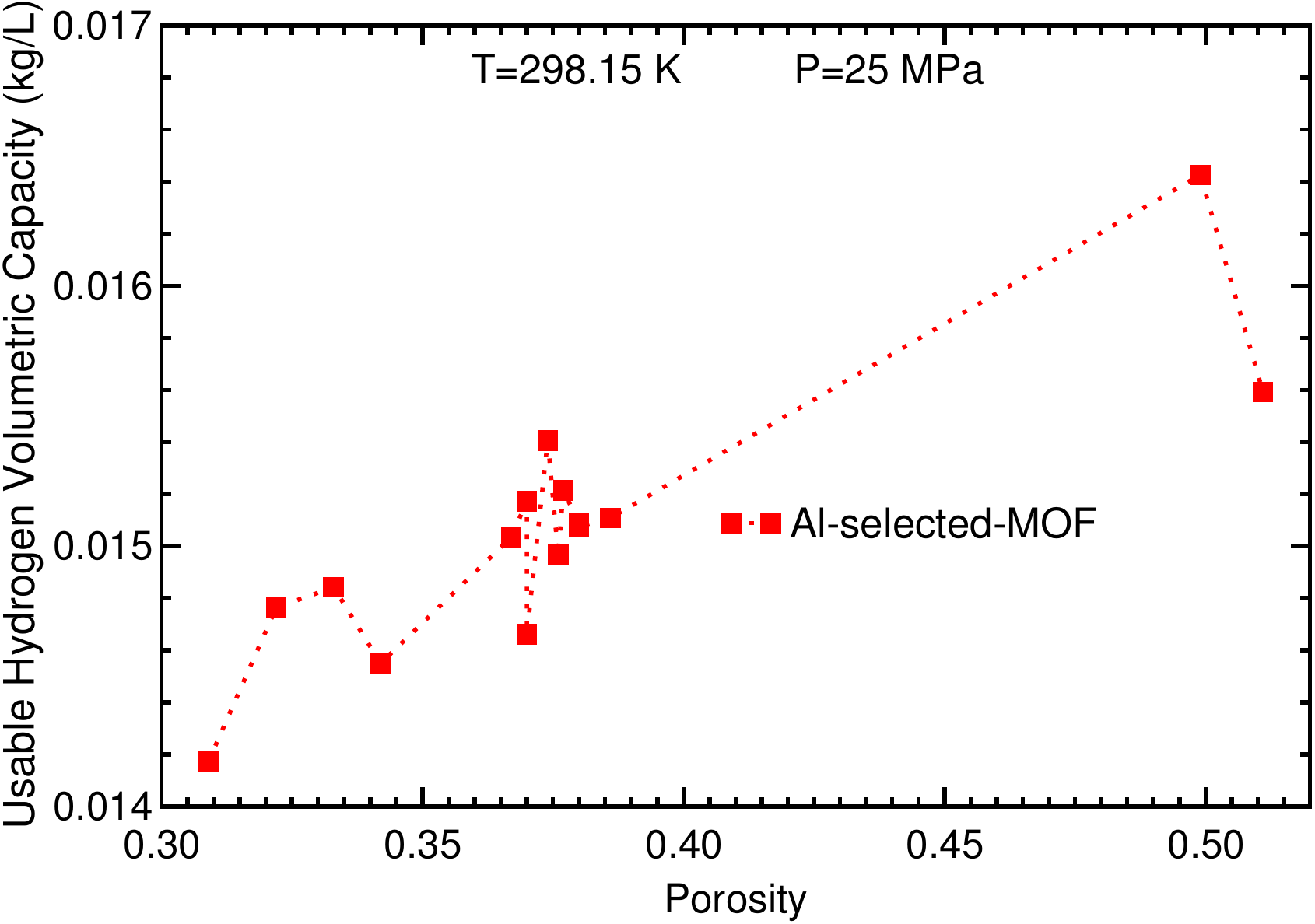}
  }
\end{center}
\vspace{-0.5cm}
\caption{Hydrogen gravimetric and volumetric usable capacity at 298.15 K and 
25 MPa vs porosity of Al-nia, the selected Al-based and the classical MOFs. 
Low panels show a zoomed-in view of the capacities of the selected Al-based 
MOFs.}
\label{h2capvsp}
\end{figure}

Plots of the gravimetric and volumetric hydrogen storage capacities of 
Al-nia, the selected Al-based and classical MOFs as a function of density 
and porosity, at 298.15 K and 25 MPa, can be found in Fig.~\ref{h2capvsd} 
and~\ref{h2capvsp}, respectively. It can be observed in those figures that 
the capacities are, in general, inversely proportional to the density and 
directly proportional to the porosity. This is a general trend: Some 
MOFs do not follow these trends. It can be also noticed in 
Fig.~\ref{h2capvsd} that the gravimetric capacities of Al-nia MOFs lie 
approximately in the general line formed by the gravimetric capacities of 
classical and selected Al-based MOFs as a function of the density, i.e., their 
gravimetric capacities have the `expected' gravimetric capacity for their 
densities. In Fig.~\ref{h2capvsp} Al-nia-MOFs present higher gravimetric 
capacity than classical and selected Al-based MOFs for similar porosities.

The selected Al-based MOFs have similar storage capacities, densities and 
porosities. Hence, in order to visualize correctly these points, specific 
plots of the capacities of this set of MOFs have been plotted and included 
in the lower panels of Fig.~\ref{h2capvsd} and~\ref{h2capvsp}.

Hydrogen storage capacities of Al-nia-MOFs are very similar (See 
Table~\ref{h2captab}). At 298.15 K and 25 MPa gravimetric and 
volumetric usable capacities were found to be 3.03 wt. \% and 0.015 
kg/L for Al-nia-MOF-1, respectively. The gravimetric and volumetric 
usable capacities of Al-nia-MOF-2 are 3.09 wt. \% and 0.016 kg/L, 
respectively. Porosity for Al-nia-MOF-1 and Al-nia-MOF-2 are 0.415 and 
0.474, respectively. More porosity means higher gravimetric and 
volumetric usable capacities.

\begin{table}[ht]
\begin{center}
\caption{\noindent
\label{h2captab}
Hydrogen volumetric (in kg/L) and gravimetric (in wt. \%) usable 
capacities at 298.15 K and 25 MPa obtained in the present GCMC 
simulations and density, $\rho$, (in kg/L) and porosity, prs., of the two novel 
Al-based MOFs, the selected Al-based and the classical MOFs.}
\bigskip
\begin{tabular}{cccccccccc}
\hline
MOF          & v$_c$  & g$_c$ & $\rho$ & prs.  & MOF     & v$_c$  & g$_c$ & $\rho$ & prs.\\
\hline
Al-nia-MOF-1 & 0.0153 & 3.03  & 0.489  & 0.415 & OYEWEW  & 0.0148 & 2.35  & 0.613  & 0.322\\
Al-nia-MOF-2 & 0.0157 & 3.09  & 0.491  & 0.474 & QOXSAA  & 0.0148 & 2.18  & 0.666  & 0.333\\
IRMOF-5      & 0.0038 & 0.14  & 2.731  & 0.030 & OTECEX  & 0.0152 & 2.51  & 0.589  & 0.370\\
IRMOF-8      & 0.0175 & 2.92  & 0.582  & 0.559 & QOXQUS  & 0.0150 & 2.49  & 0.589  & 0.367\\
IRMOF-12     & 0.0162 & 3.71  & 0.420  & 0.585 & OTEDOI  & 0.0151 & 2.51  & 0.585  & 0.380\\
IRMOF-14     & 0.0156 & 3.98  & 0.376  & 0.582 & QOXVEH  & 0.0152 & 2.54  & 0.584  & 0.377\\
IRMOF-15     & 0.0162 & 4.38  & 0.354  & 0.861 & QOXSUU  & 0.0146 & 2.45  & 0.578  & 0.342\\
IRMOF-20     & 0.0187 & 2.78  & 0.655  & 0.535 & RIXPIZ  & 0.0150 & 2.56  & 0.570  & 0.376\\
MUZKEZ       & 0.0156 & 3.70  & 0.406  & 0.511 & OTEDUO  & 0.0147 & 2.53  & 0.565  & 0.370\\
GUFYUE       & 0.0164 & 2.62  & 0.609  & 0.499 & WOFQEQ  & 0.0154 & 2.65  & 0.565  & 0.374\\
QOXTOP       & 0.0142 & 2.07  & 0.669  & 0.309 & QOXRIH  & 0.0151 & 2.61  & 0.563  & 0.386\\
             &        &       &        &       & QOXRAZ  & 0.0151 & 2.64  & 0.555  & 0.380\\
\hline
\end{tabular}
\end{center}
\end{table}

MOFs with the highest and the lowest gravimetric and volumetric 
capacities at 25 MPa of some selected Al-based and classical MOFs have 
been chosen to compare with the Al-nia-MOFs (See Table~\ref{h2captab}). 
Among selected Al-based MOFs, the one with the best volumetric capacity 
at 25 MPa, 0.016 kg/L, has the CCDC database identifier GUFYUE and 
the one with the best gravimetric capacity, 3.7 wt. \%, is Al-soc-MOF-1 
(CCDC database identifier MUZKEZ). This last MOF is reported as the 
current record holder for \ch{CO2} storage \cite{alniamof2022,alsocmof15}. 
On the other hand, the MOF with the CCDC database 
identifier QOXTOP has the lowest gravimetric and volumetric capacities, 
2.1 wt. \% and 0.014 kg/L, respectively. Among the classical MOF 
collection, IRMOF-20 has the highest volumetric capacity (0.019 kg/L), 
IRMOF-15 has the highest gravimetric capacity, 4.4 wt. \%. IRMOF-5 has 
the lowest gravimetric and volumetric capacities, 0.14 wt. \% and 0.004 
kg/L, respectively, of all the MOFs in Table~\ref{h2captab}.

\begin{figure}[htb]
\vspace{0.0cm}
\begin{center}
  \leavevmode
  \put(-220,   0){%
	\includegraphics[width=7cm]{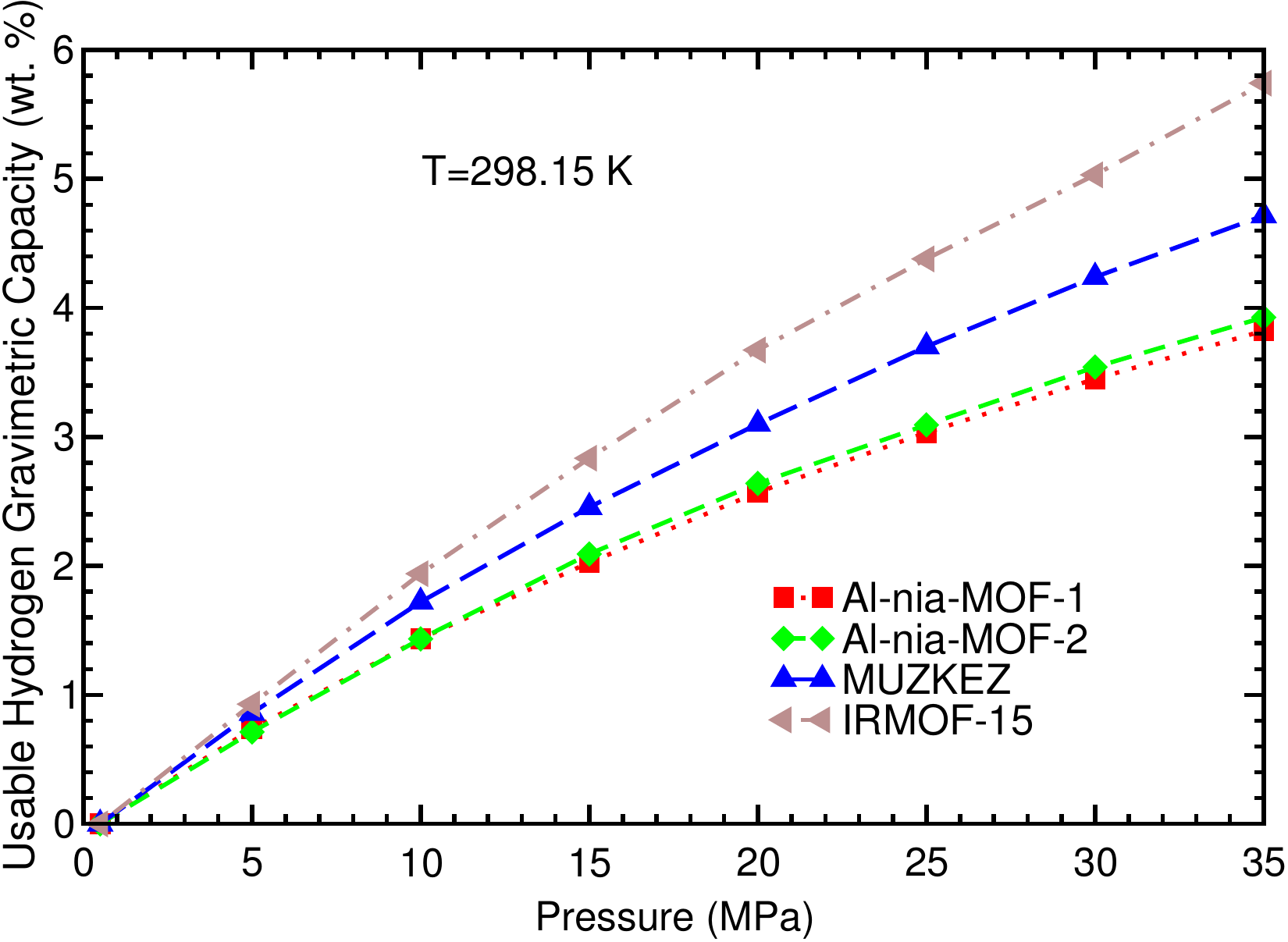}
  }
  \put(   0,   0){%
	\includegraphics[width=7.1cm]{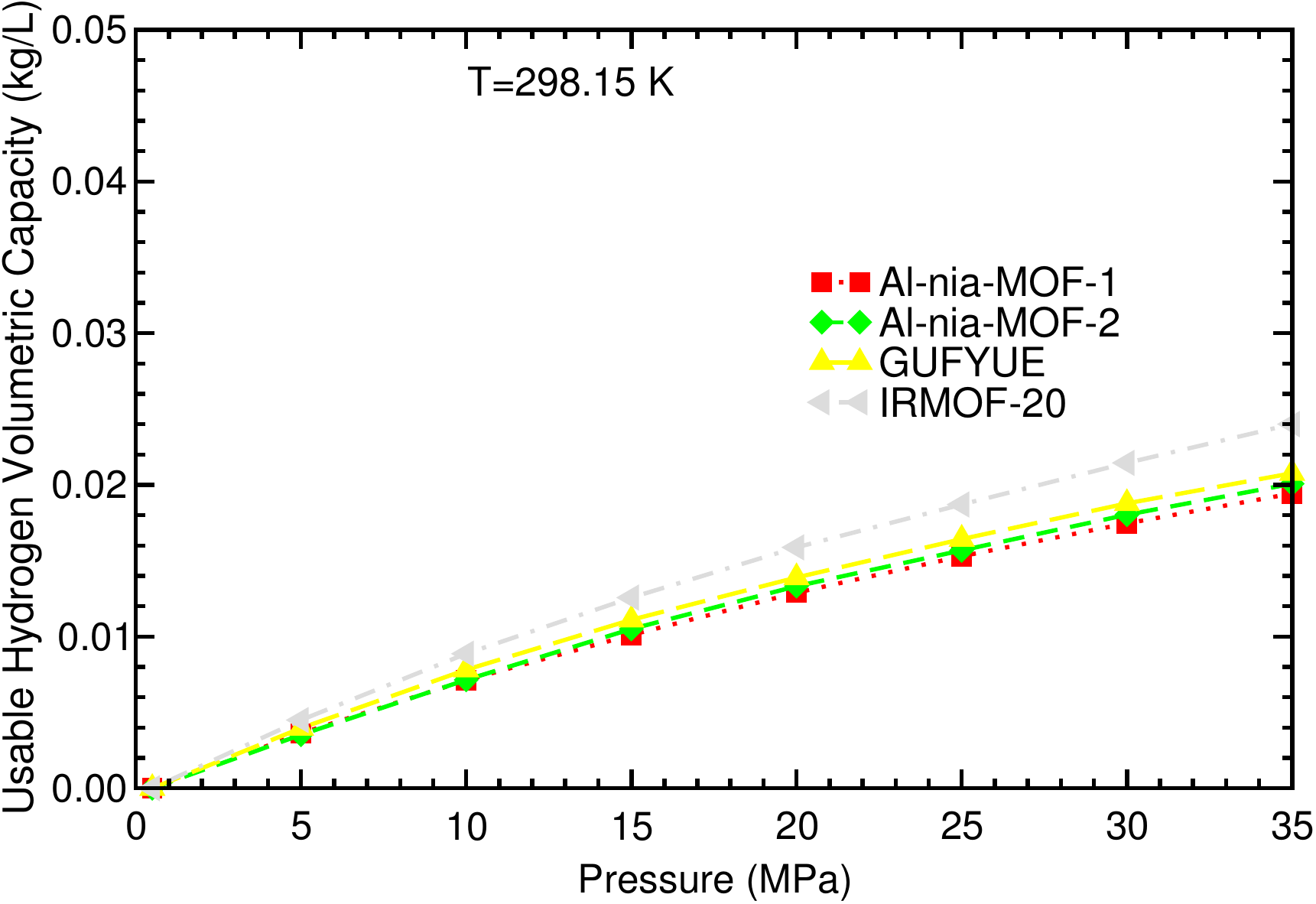}
  }
\end{center}
\vspace{-0.5cm}
\caption{Hydrogen volumetric and gravimetric usable capacities vs 
pressure at room temperature of some selected MOFs and the two novel 
Al-nia MOFs.}
\label{h2cap}
\end{figure}

\begin{figure}[htb]
\vspace{0.0cm}
\begin{center}
\leavevmode
\put(-220,   0){%
\includegraphics[width=7cm]{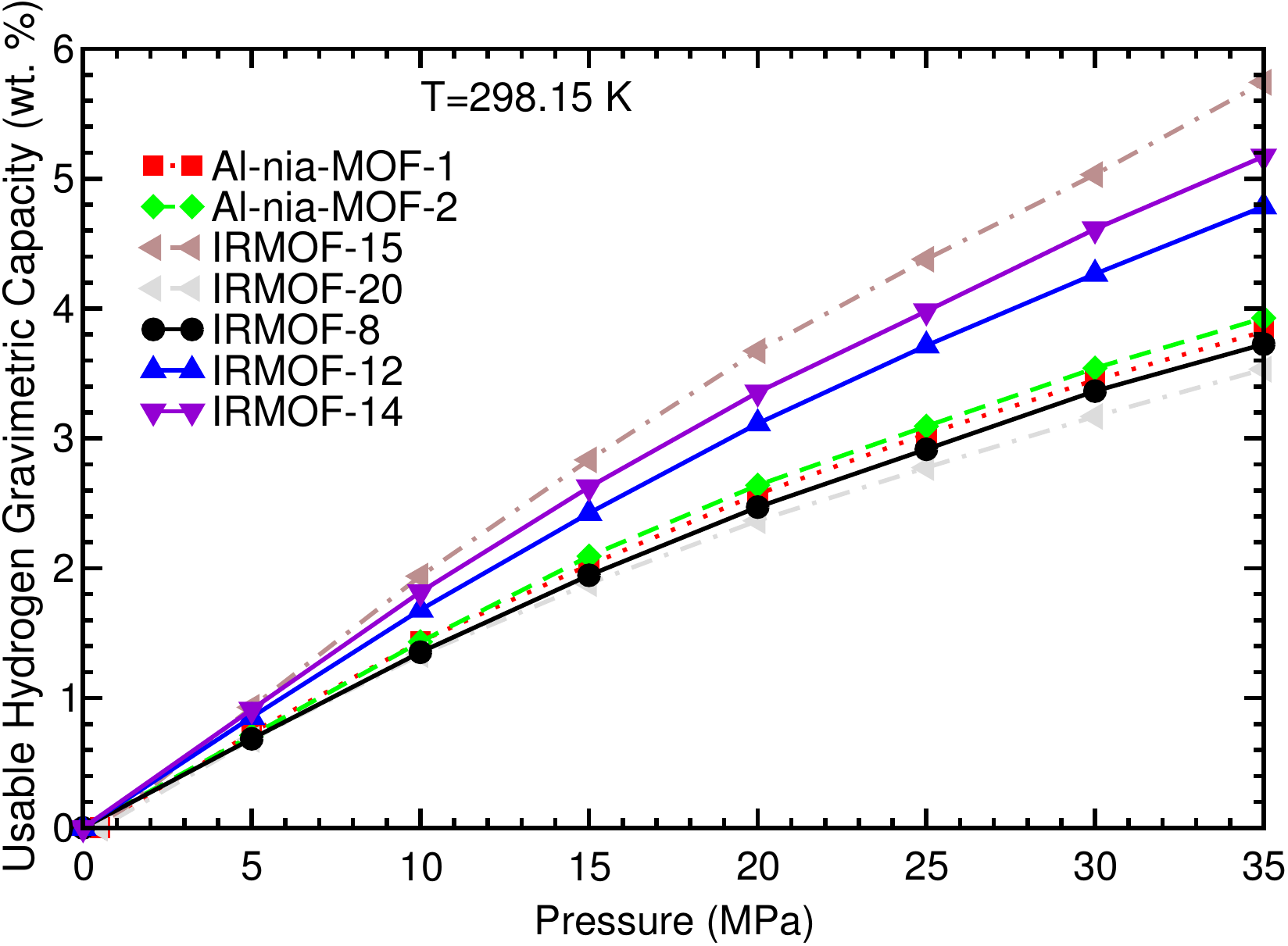}
}
\put(   0,   0){%
\includegraphics[width=7.1cm]{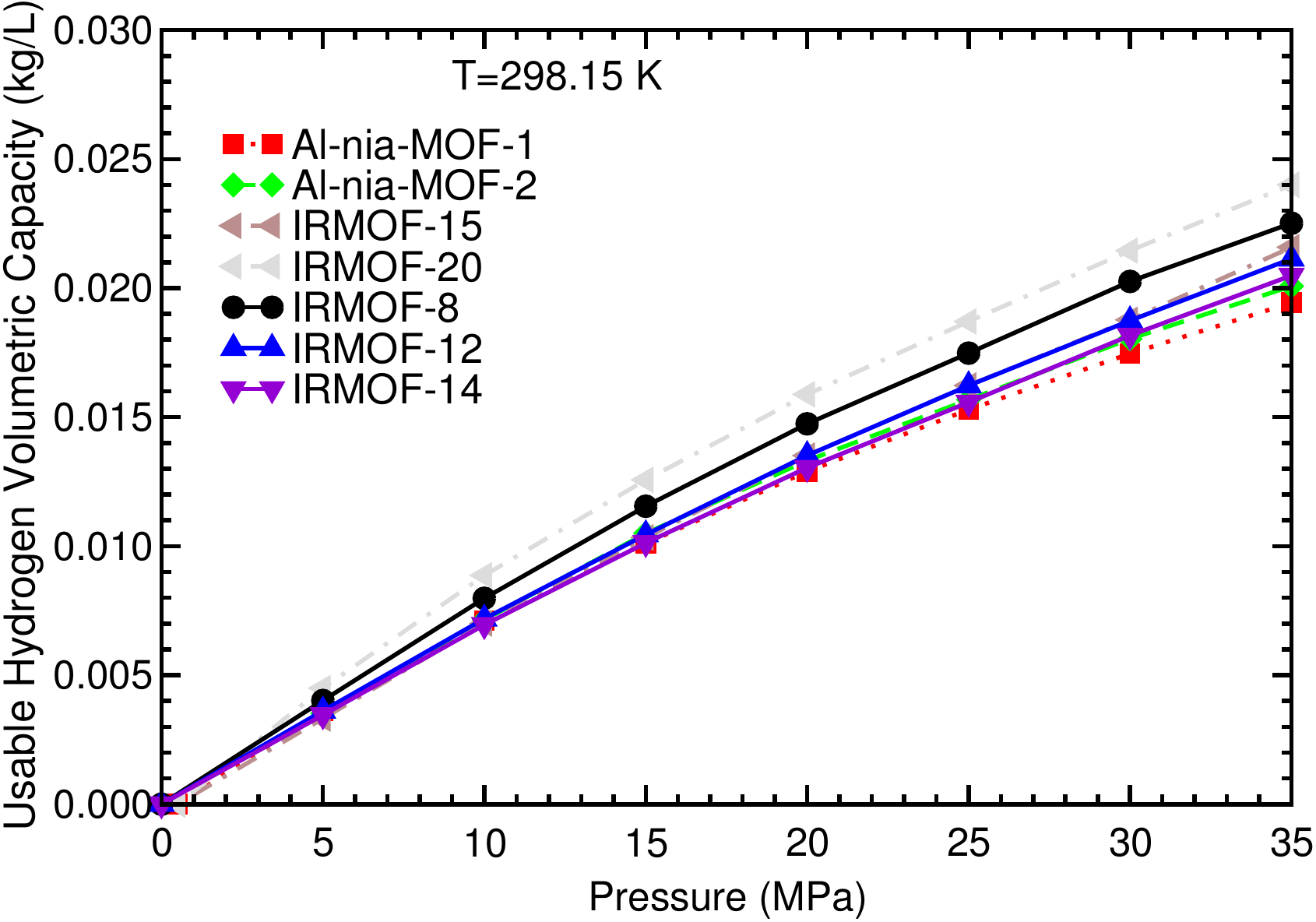}
}
\end{center}
\vspace{-0.5cm}
\caption{Hydrogen volumetric and gravimetric usable capacities vs 
pressure at room temperature of some selected classical MOFs and the two novel 
Al-nia MOFs.}
\label{classics}
\end{figure}

GCMC simulations of the \ch{H2} storage capacities of 
MUZKEZ, GUFYUE, IRMOF-8, 12, 14, 15 and 20, and of the two novel Al-nia MOFs 
have been carried out at 298.15 K and pressures between 0.5 and 35 MPa. 
The gravimetric and volumetric usable capacities are compared on 
Fig.~\ref{h2cap} and~\ref{classics}, as a function of the pressure. Gravimetric capacities 
are quite different. The gravimetric capacities of all these MOFs 
increase slowly and approximately linearly with the pressure. IRMOF-15 
has the largest gravimetric capacity at any pressure and room 
temperature. MUZKEZ MOF gravimetric capacity is in between the 
gravimetric capacities of IRMOF-15 and Al-nia MOFs, which have lower 
values.

The volumetric capacities are very similar. Volumetric capacities for 
Al-nia, GUFYUE and QOXTOP MOFs are slightly smaller than the volumetric 
capacity of IRMOF-15 and IRMOF-20. The volumetric capacity of IRMOF-5 
is much lower than the volumetric capacities of the other MOFs. From 
the highest to the lowest, the order is IRMOF-15, IRMOF-20, GUFYUE, 
Al-nia-MOF-2, Al-nia-MOF-1, QOXTOP and IRMOF-5.

The gravimetric storage capacities of the two novel Al-nia MOFs are 
high at 25-35 MPa, because their densities are low, below the water 
density, 1 kg/L. These MOFs are made of light elements (C, O, H and N) 
and of one of the lightest metals, Aluminum, and their structure is 
very porous (their hydrogen porosity is about 0.5). This explains their 
low density, below the water density, 1 kg/L, which, in turn, explains 
their high gravimetric storage capacities. As regards their volumetric 
storage capacities, they are high due to their high porosity, about 0.5.

With the aim of exploring the gravimetric and volumetric hydrogen storage 
capacities, a comparative study was carried out involving the two novel 
Al-nia-MOFs and a selection of the classical MOFs with the highest 
capacities: IRMOF-8, 12, 14, 15, and 20. GCMC simulations were performed 
at 298.15 K and pressures ranging from 0.5 to 35 MPa. The results of these 
simulations have been illustrated in Fig.~\ref{classics}.

The gravimetric capacities of the two Al-nia-MOFs, IRMOF-8 and 20 are very 
similar, while the gravimetric capacities of the other MOFs (IRMOF-12, 14 
and 15) in Fig.~\ref{classics} are about 20-50 \% higher. This second group 
of MOFs has lower densities than the first group (See Table~\ref{h2captab}). 
This explains the higher gravimetric capacities of the second group of MOFs.
The volumetric capacities of IRMOF-12, 14, and 15 and Al-nia-MOFs are very 
similar. The volumetric capacities of IRMOF-8 and 20 are slightly higher 
than the capacities of the other MOFs.

\begin{figure}[htb]
\vspace{0.0cm}
\begin{center}
\leavevmode
\put(-220, 150){%
\includegraphics[width=7cm]{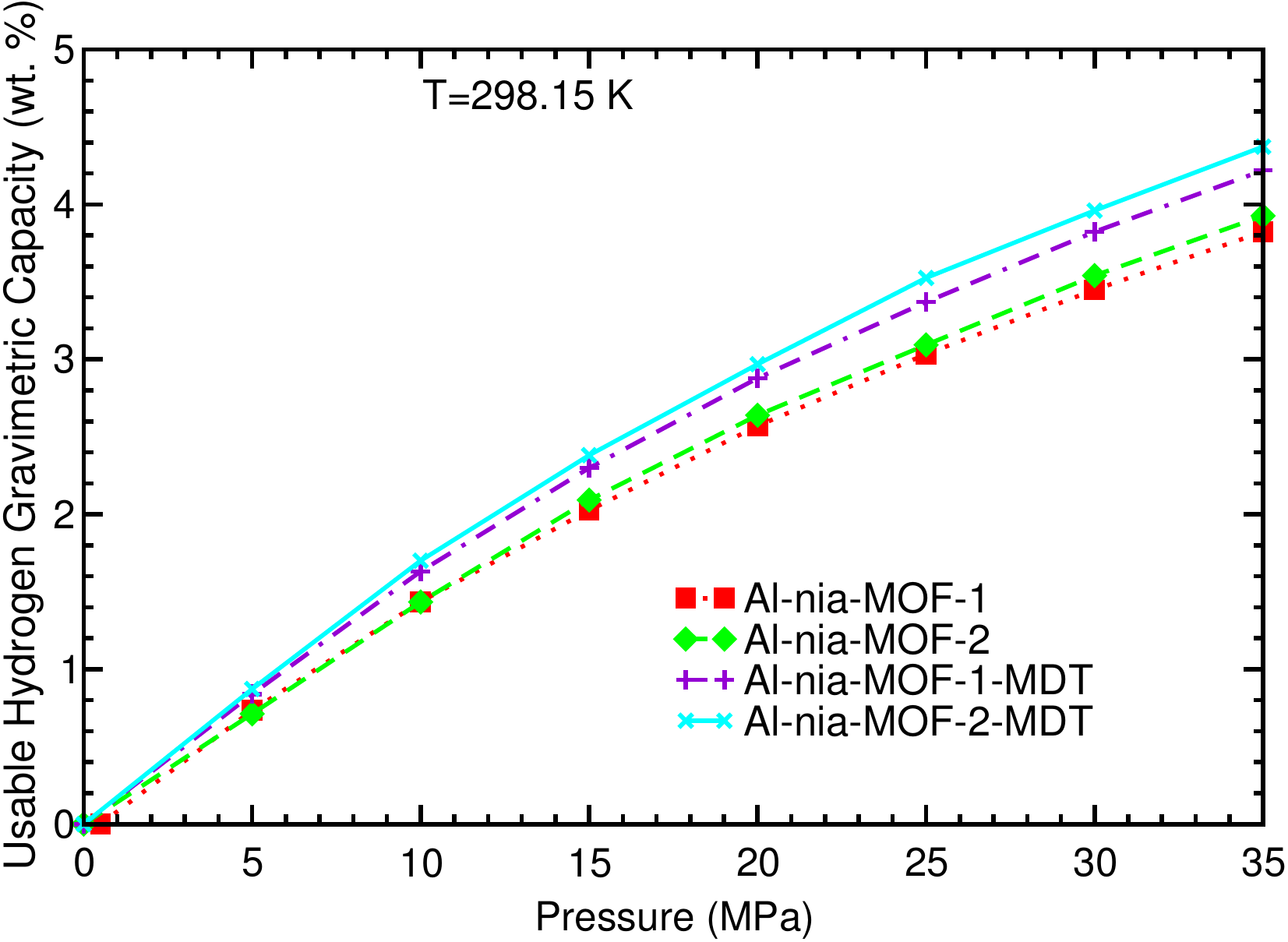}
}
\put(   0, 150){%
\includegraphics[width=7cm]{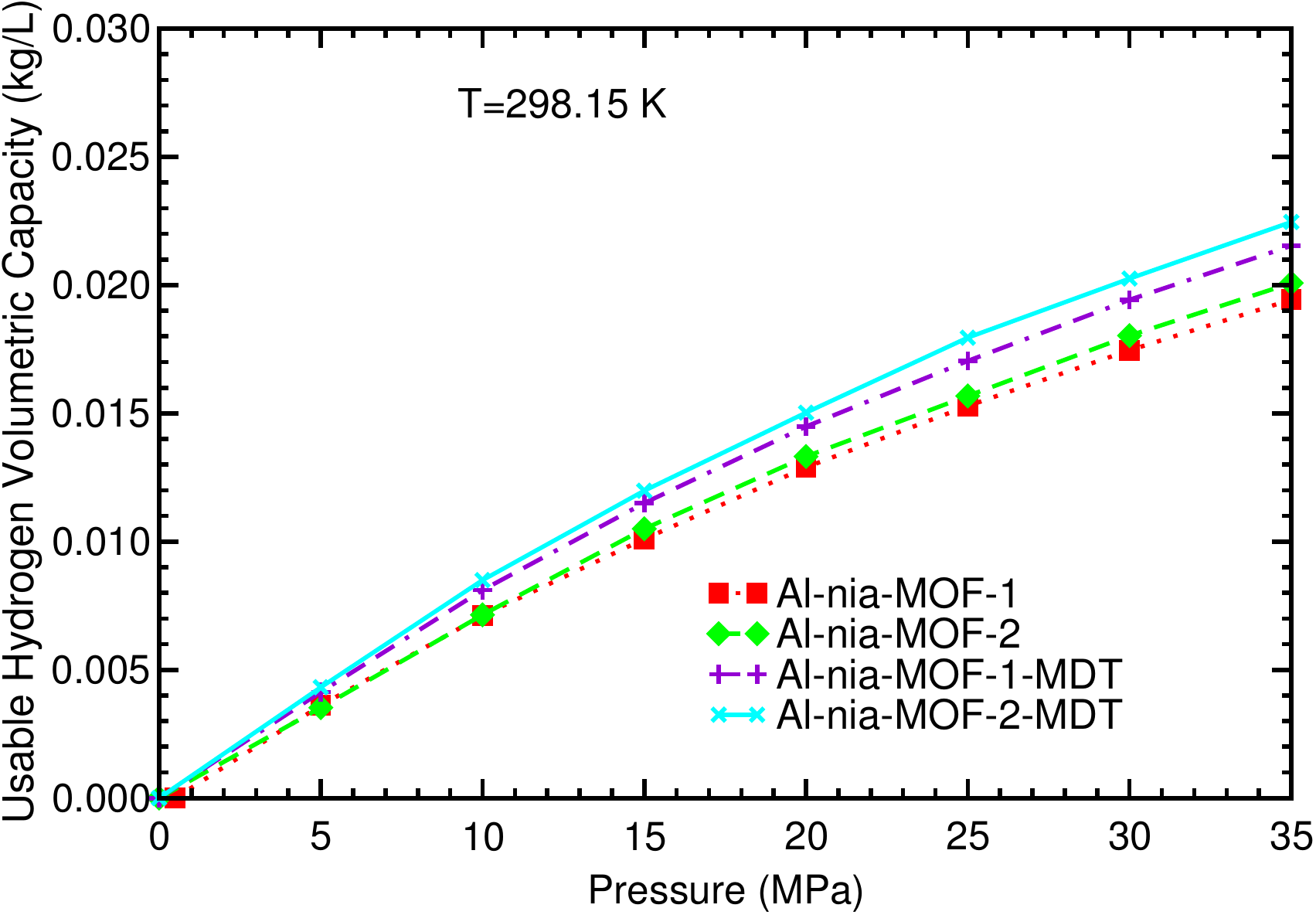}
}
\put(-220,   0){%
\includegraphics[width=7cm]{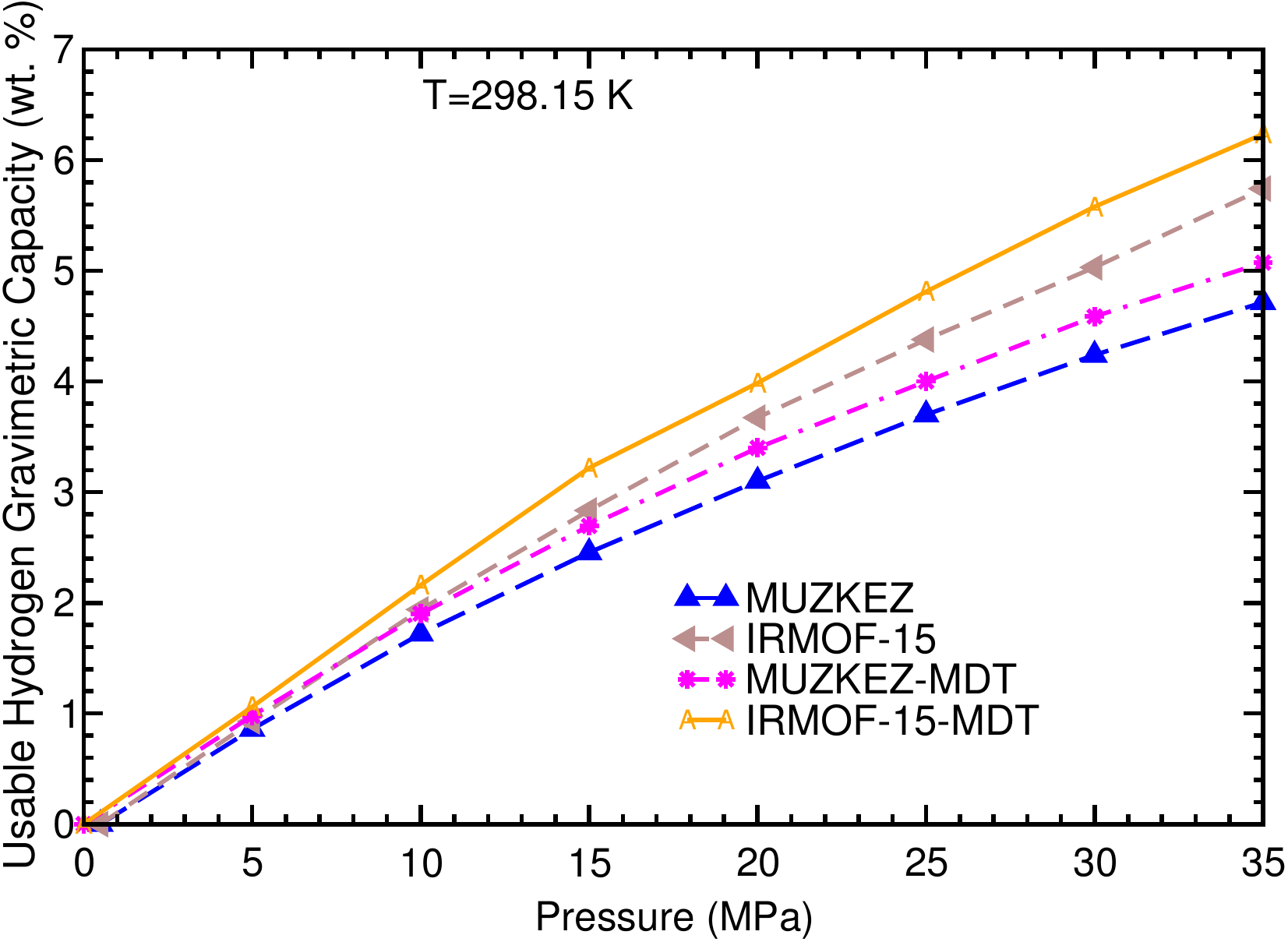}
}
\put(   0,   0){%
\includegraphics[width=7cm]{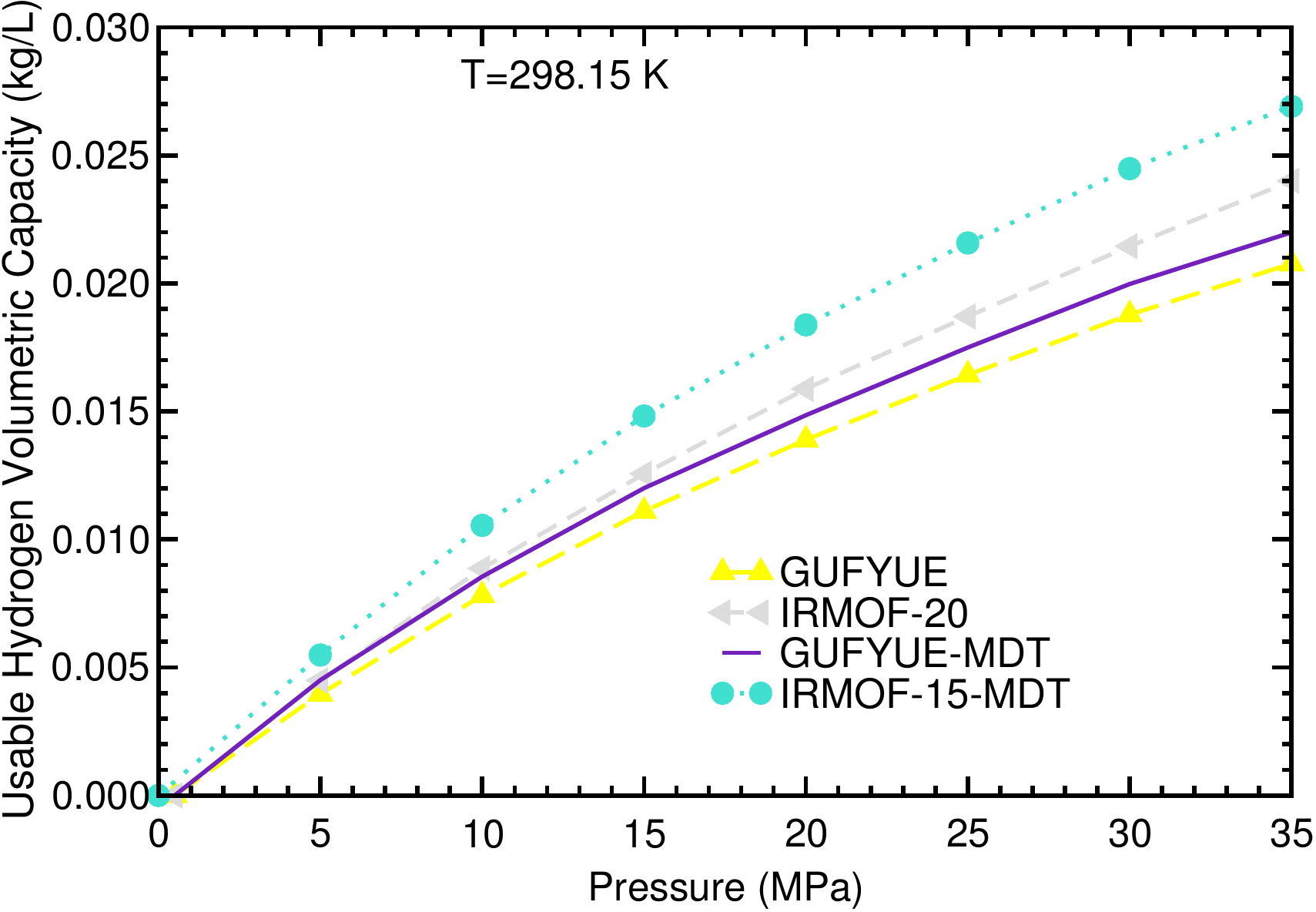}
}
\end{center}
\vspace{-0.5cm}
\caption{
Hydrogen gravimetric and volumetric usable capacity vs pressure at room 
temperature of the Al-nia-MOFs (upper panels) and the best Al-based MOF, 
MUZKEZ, and the best classical MOF for gravimetric, IRMOF-15, and for 
volumetric capacity, IRMOF-20 (lower panels).}
\label{h2MDTset}
\end{figure}

GCMC simulations of the hydrogen storage capacities of some MOFs with a 
different set of LJ coefficients, the MDT set, were carried out (See 
Table~\ref{coeffsLJMDT}). The storage capacities at room temperature of 
Al-nia-MOFs and Al-nia-MOFs-MDT are compared in Fig.~\ref{h2MDTset} as 
functions of pressure. Remarkably, the capacities of Al-nia-MOFs and 
Al-nia-MOFs-MDT display a high degree of similarity. The capacities of the 
Al-nia-MOFs-MDT are slightly higher than those of the Al-nia-MOFs. The 
differences between the capacities of Al-nia-MOF-1 and Al-nia-MOF-2 are 
almost identical to the differences between the capacities of 
Al-nia-MOF-1-MDT and Al-nia-MOF-2-MDT. The trend observed in the 
Al-nia-MOF-MDT curves closely resembled that of the Al-nia-MOF curves.

The storage capacities at 298.15 K of the best Al-based and classic MOFs 
with the original and MDT LJ sets are compared in Fig.~\ref{h2MDTset} as 
functions of the pressure. The capacities obtained with these two LJ sets 
are, in general, similar. The capacities of IRMOF-15 and 20 obtained 
with the MDT LJ set are about 7-12 \% higher than the capacities obtained 
with the original LJ set of coefficients. These differences between the 
results obtained with the two LJ sets are mainly due to 
the fact that the MDL LJ $\epsilon$ coefficient of the hydrogen molecule, 
0.003163 eV, is slightly larger than the counterpart in the original LJ 
set, 0.002870 eV (See Tables~\ref{coeffsLJ} and~\ref{coeffsLJMDT}).

\subsection{Methane Storage Capacities}

The methane storage capacities have been also investigated. The two 
novel Al-nia MOFs have similar methane storage capacities. At 298.15 K 
and 25 MPa the usable methane gravimetric and volumetric capacities 
were found to be, respectively, 27.2 wt. \% and 0.183 kg/L for 
Al-nia-MOF-1, and 28.1 wt. \% and 0.192 kg/L for the Al-nia-MOF-2. 
These capacities are high and very close to the DOE methane targets.

\begin{figure}[htb]
\vspace{0.0cm}
\begin{center}
  \leavevmode
  \put(-220, 150){%
	\includegraphics[width=7cm]{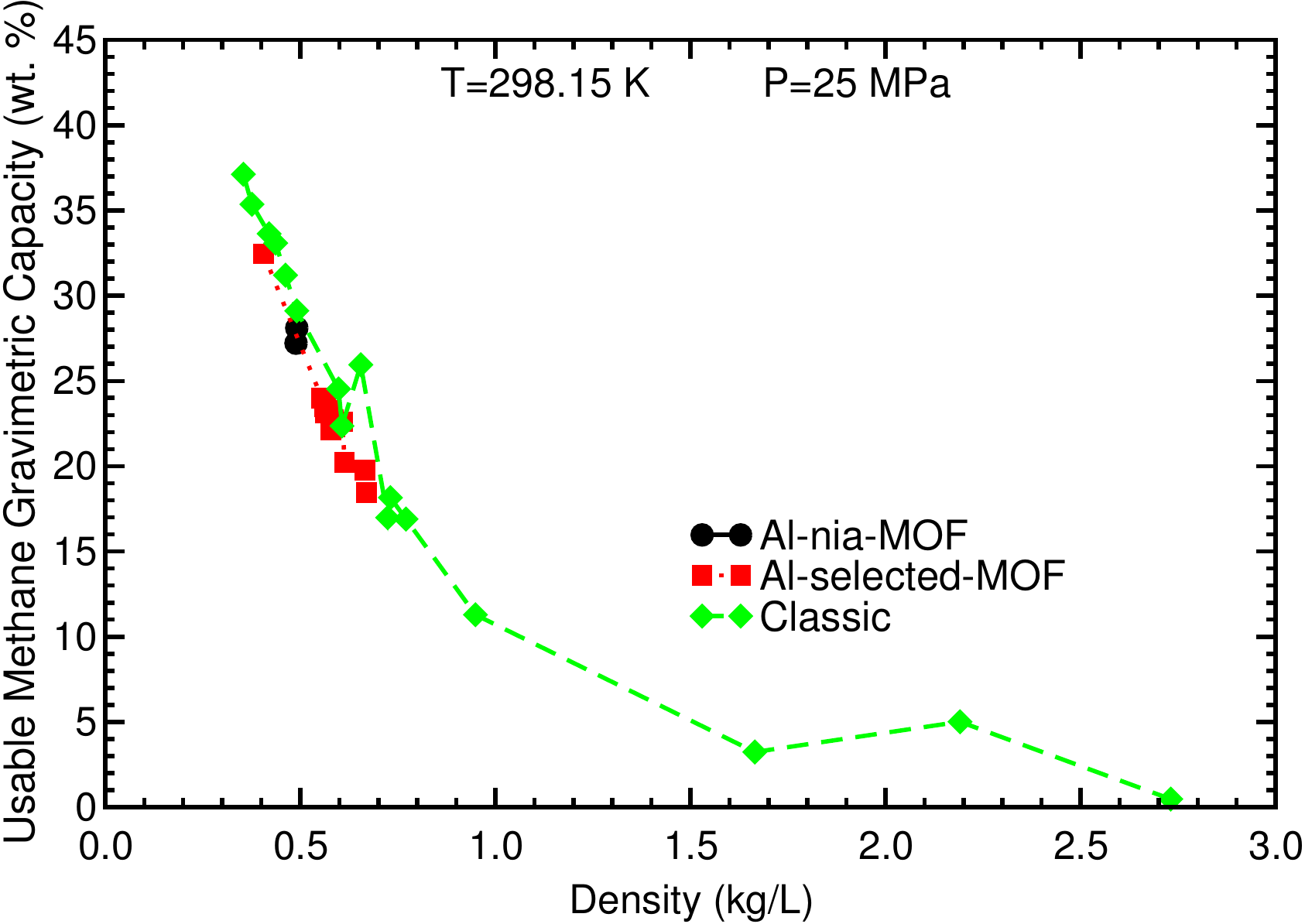}
  }
  \put(   0, 150){%
	\includegraphics[width=7cm]{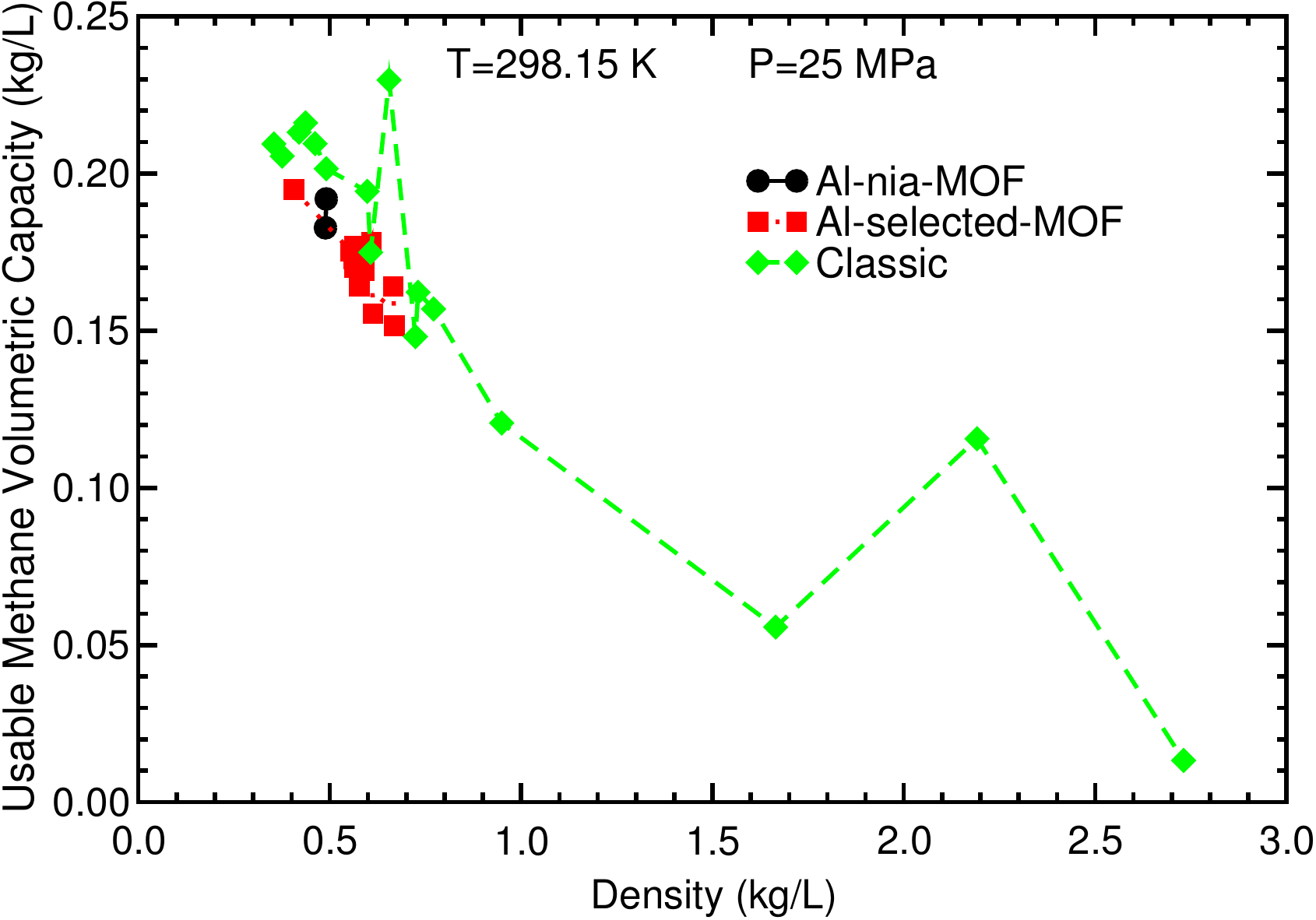}
  }
  \put(-220,   0){%
	\includegraphics[width=7cm]{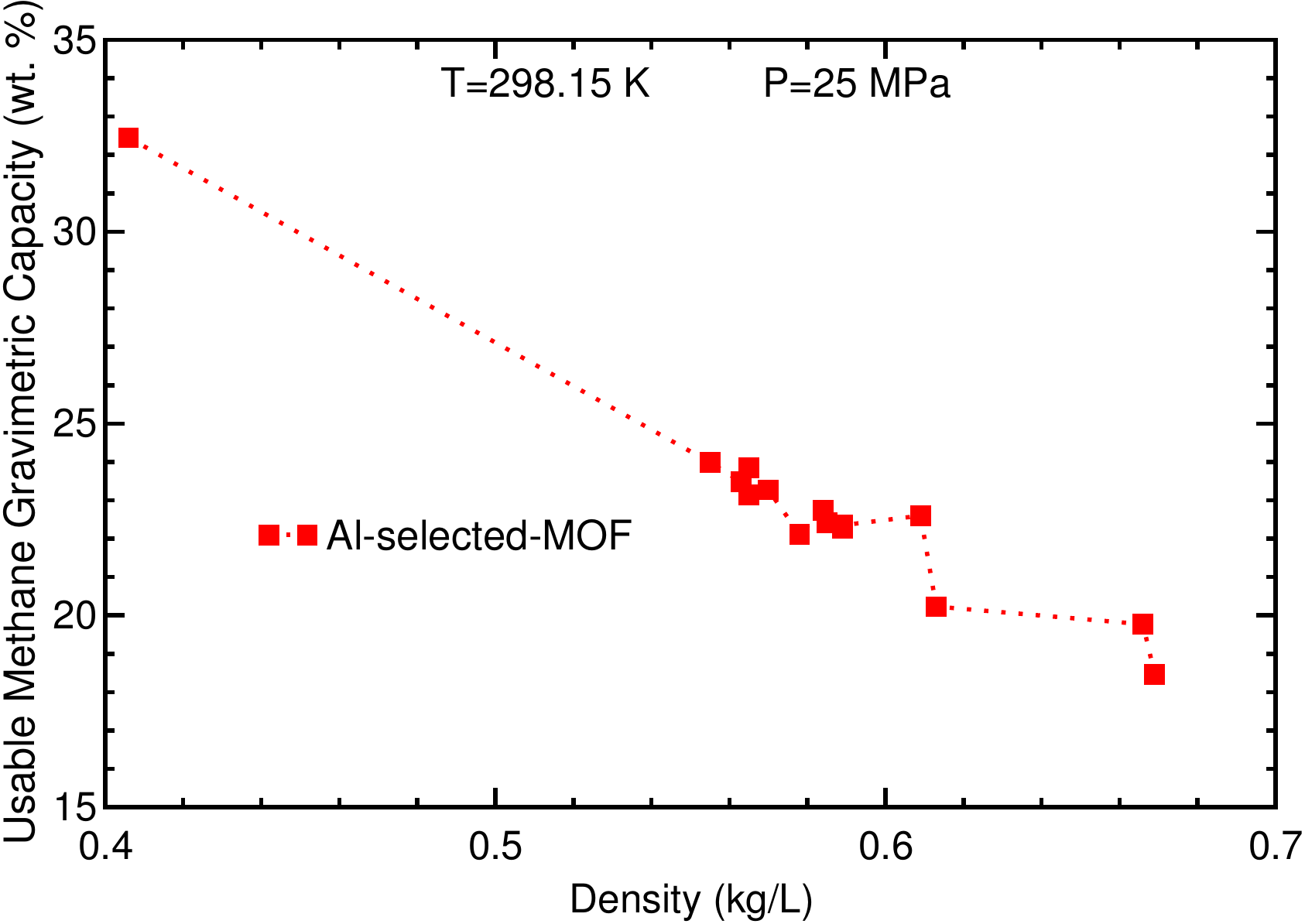}
  }
  \put(   0,   0){%
	\includegraphics[width=7cm]{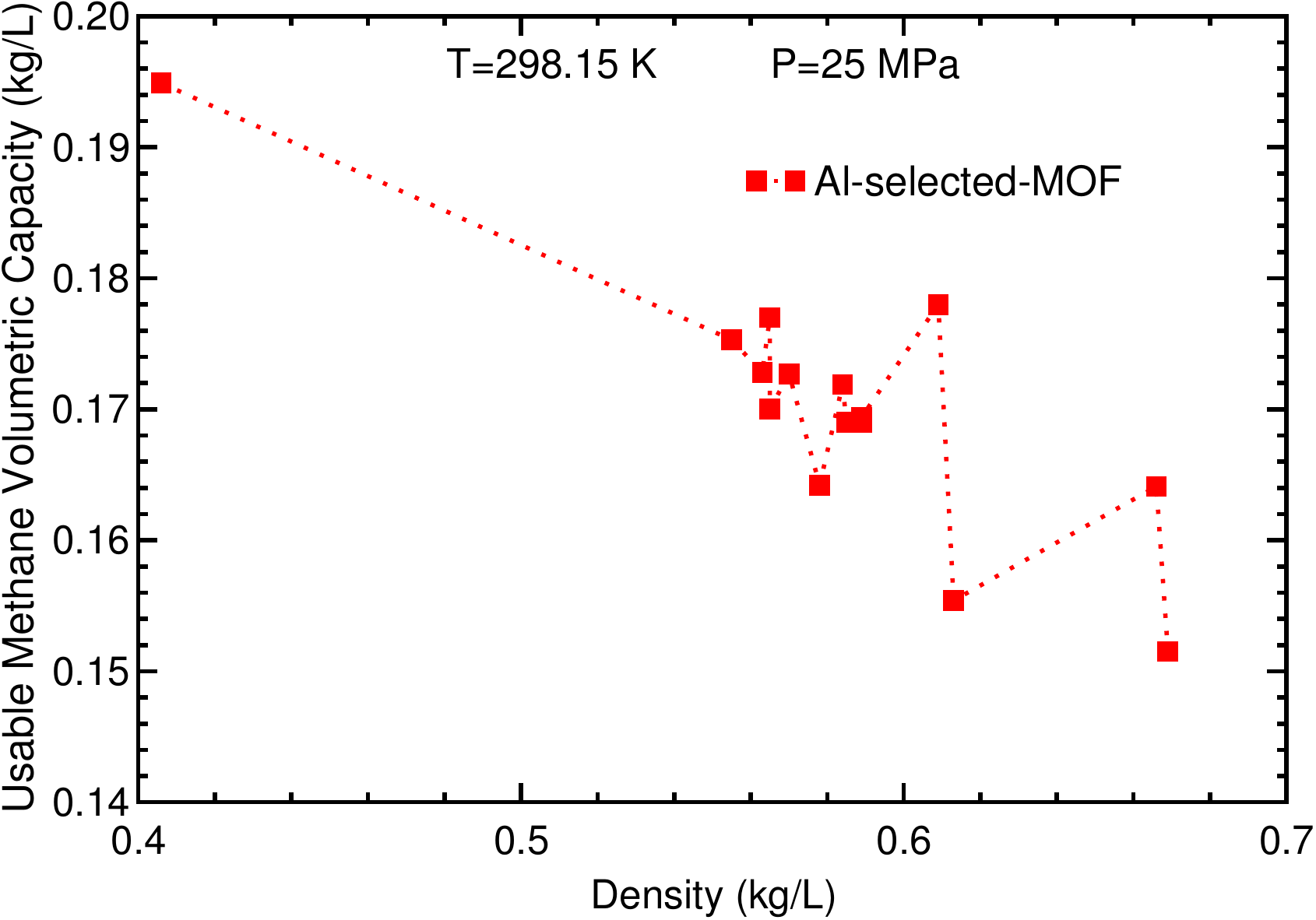}
  }
\end{center}
\vspace{-0.5cm}
\caption{Methane gravimetric and volumetric usable capacity at 298.15 K and 
25 MPa vs density of Al-nia, the selected Al-based and the classical MOFs. 
Low panels show a zoomed-in view of the selected Al-based MOFs.}
\label{ch4capvsd}
\end{figure}

\begin{figure}[htb]
\vspace{0.0cm}
\begin{center}
  \leavevmode
  \put(-220, 150){%
	\includegraphics[width=7cm]{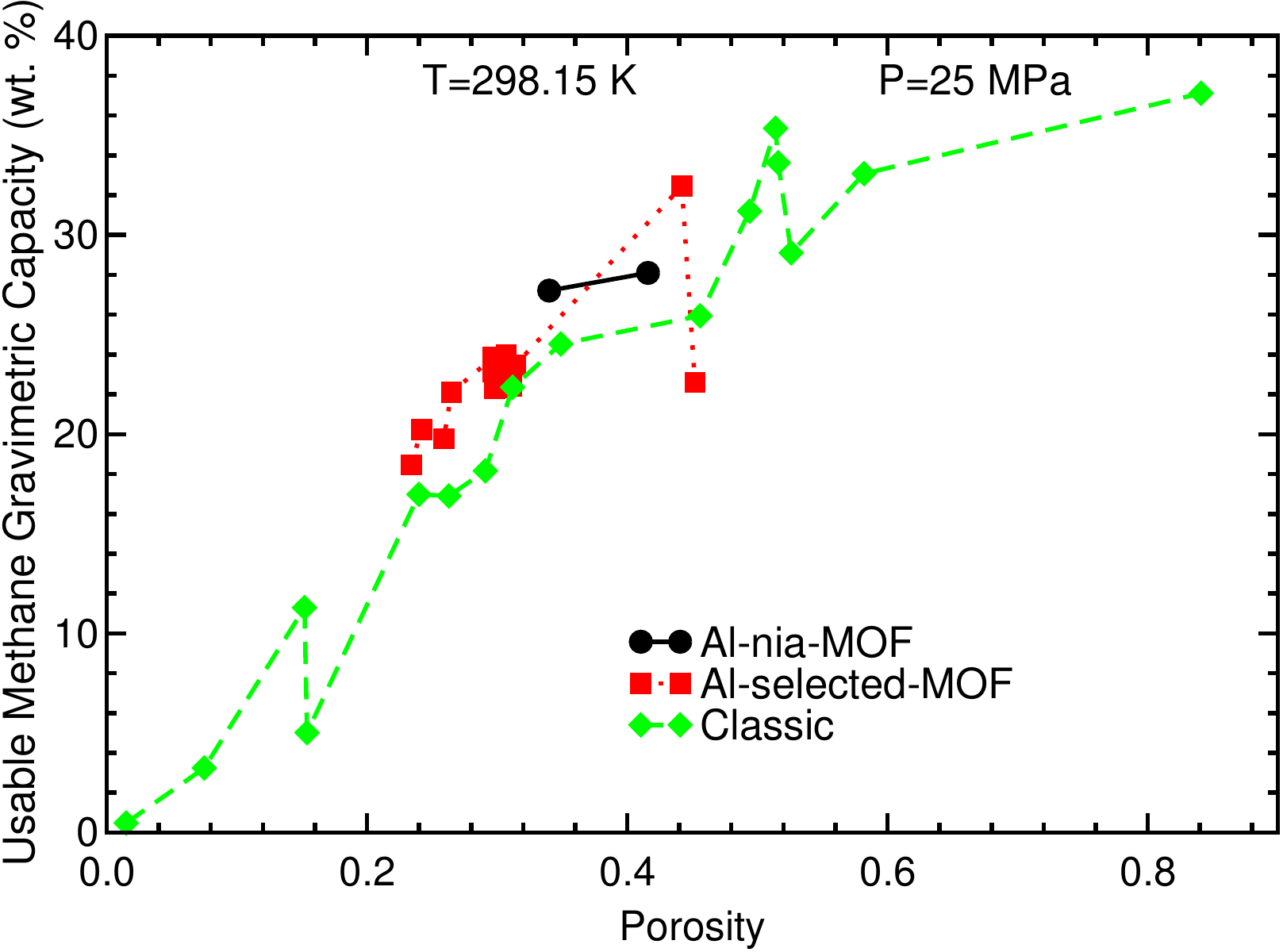}
  }
  \put(   0, 150){%
	\includegraphics[width=7cm]{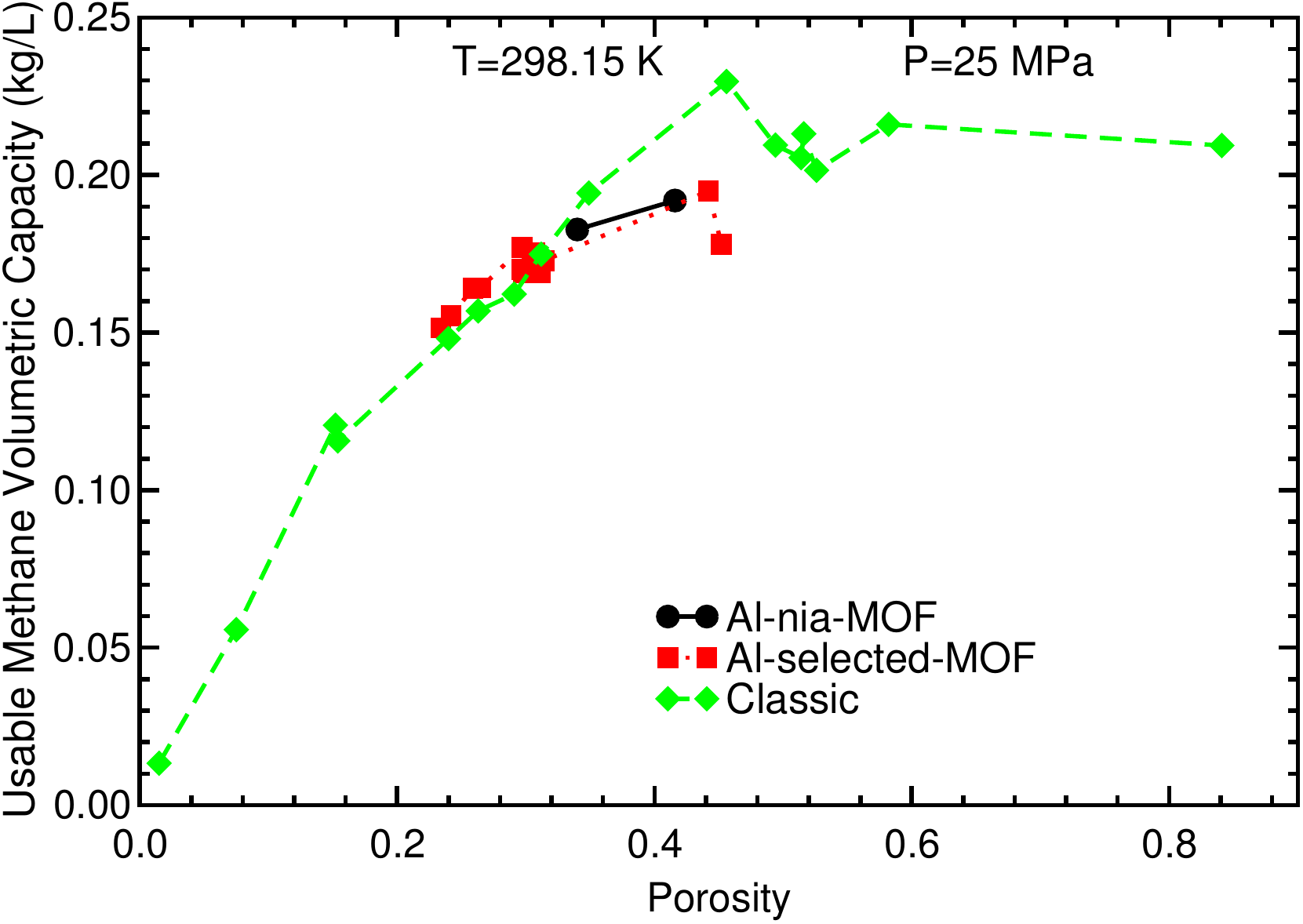}
  }
  \put(-220,   0){%
	\includegraphics[width=7cm]{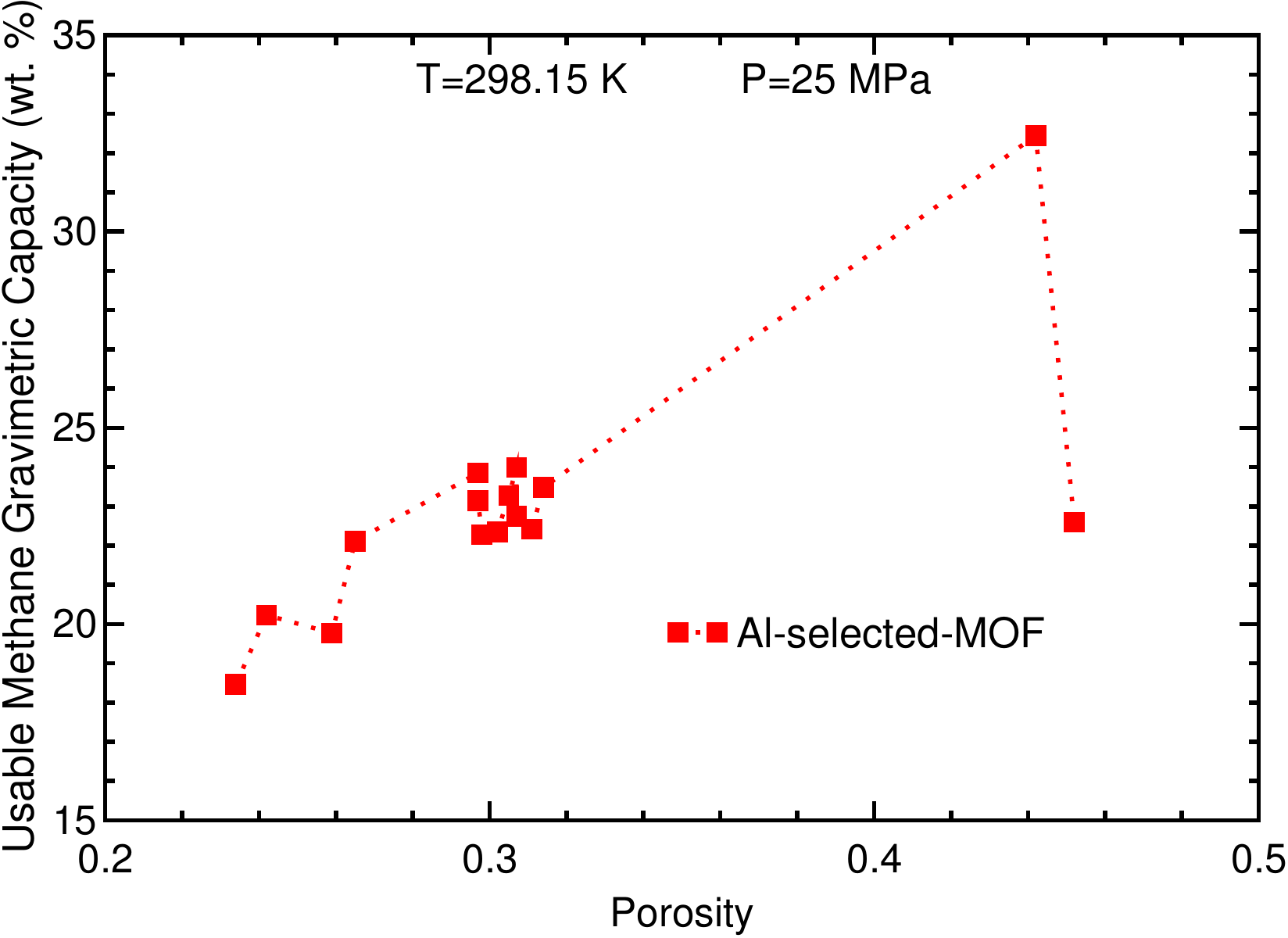}
  }
  \put(   0,   0){%
	\includegraphics[width=7cm]{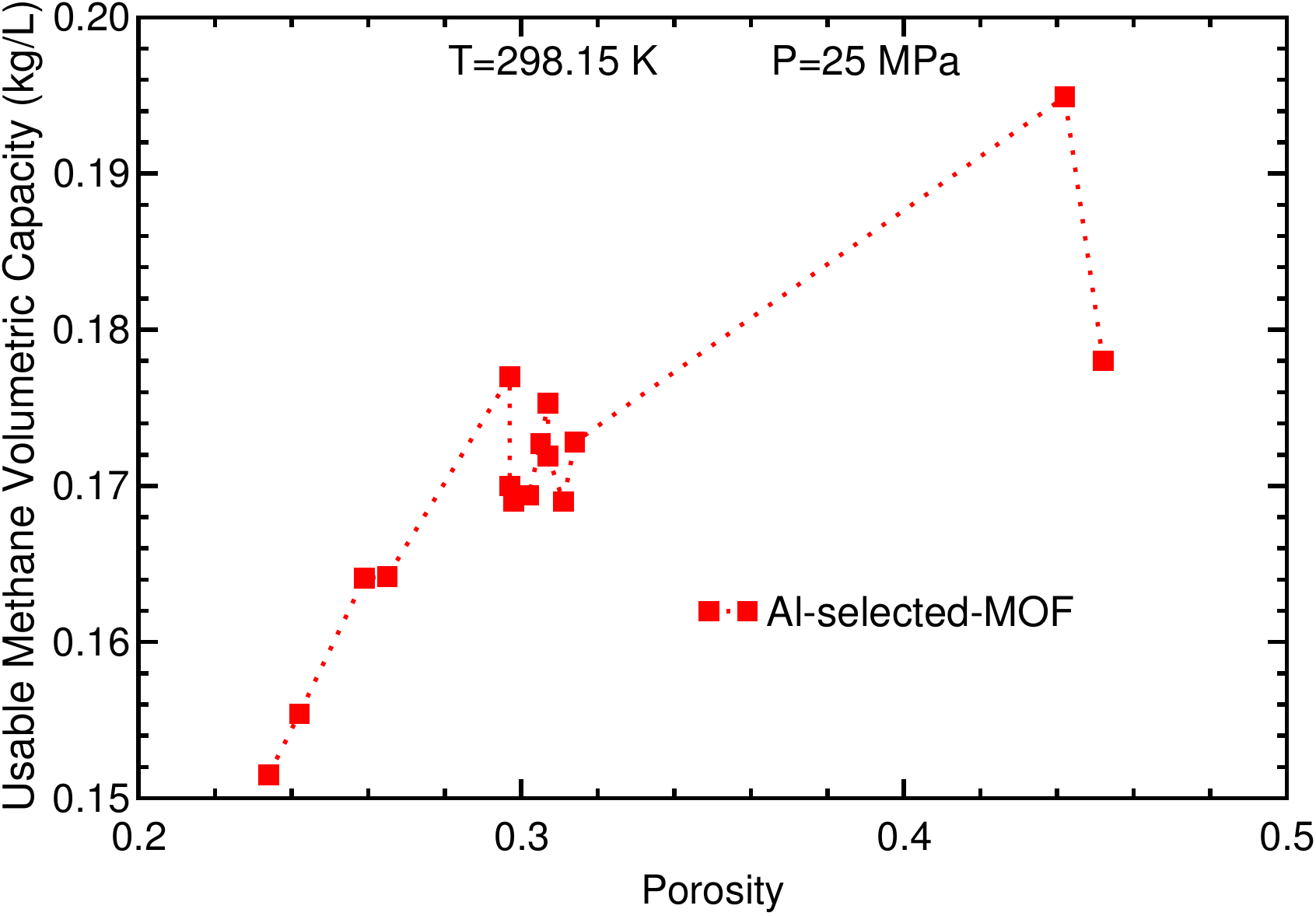}
  }
\end{center}
\vspace{-0.5cm}
\caption{Methane gravimetric and volumetric usable capacity at 298.15 K and 
25 MPa vs porosity of Al-nia, the selected Al-based and the classical MOFs. 
Low panels show a zoomed-in view of the selected Al-based MOFs.}
\label{ch4capvsp}
\end{figure}

The methane storage capacities of Al-nia, some selected Al-based and 
classical MOFs, at 298.15 and 25 MPa, as functions of density and porosity 
have been plotted in Fig.~\ref{ch4capvsd} and~\ref{ch4capvsp}, respectively. 
The methane storage capacities, densities and porosities of the selected 
Al-based MOFs are similar and therefore, to visualize them correctly, the 
results of these MOFs have been represented in the lower panels of those 
figures. The storage capacities are, in general, inversely proportional to 
the density and directly proportional to the porosity. The gravimetric 
methane capacities of the two novel Al-nia MOFs lie approximately in the 
general line formed by the gravimetric capacities of classical and selected 
Al-based MOFs as a function of the density. The methane and hydrogen storage 
capacities share these same trends or dependences on the density and 
porosity.

As it was done for hydrogen storage, MOFs with the highest gravimetric and 
volumetric methane storage capacities at 25 MPa from the selected Al-based 
and the classical MOFs were chosen (See Table~\ref{ch4captab}) to make GCMC 
simulations between 0.5 and 35 MPa. Among the selected Al-based MOFs, the 
one with the best methane volumetric and gravimetric capacities, 0.195 kg/L 
and 32.5 wt. \%, is named MUZKEZ, and it is also one of the selected 
Al-based MOFs of the CCDC database with the best hydrogen gravimetric 
capacity. The one with the lowest methane volumetric and gravimetric 
capacities, 0.151 kg/L and 18.5 wt. \%, respectively, is QOXTOP, again in 
agreement with the hydrogen capacities results.

\begin{table}[ht]
\begin{center}
\caption{\noindent
\label{ch4captab}
Methane volumetric (in kg/L) and gravimetric (in wt. \%) usable capacities 
at 298.15 K and 25 MPa obtained in the present GCMC simulations and density, 
$\rho$, (in kg/L) and porosity (prs.) of the two novel Al-based MOFs, the 
selected Al-based and the classical MOFs.}
\bigskip
\begin{tabular}{cccccccccc}
\hline
MOF          & v$_c$ & g$_c$ & $\rho$ & prs.  & MOF     & v$_c$ & g$_c$ & $\rho$ & prs.\\
\hline
Al-nia-MOF-1 & 0.183 & 27.2  & 0.489  & 0.340 & OYEWEW  & 0.155 & 20.2  & 0.613  & 0.242\\
Al-nia-MOF-2 & 0.192 & 28.1  & 0.491  & 0.416 & QOXSAA  & 0.164 & 19.8  & 0.666  & 0.259\\
IRMOF-5      & 0.013 &  0.5  & 2.731  & 0.015 & OTECEX  & 0.169 & 22.4  & 0.589  & 0.302\\
IRMOF-8      & 0.223 & 27.7  & 0.582  & 0.491 & QOXQUS  & 0.169 & 22.3  & 0.589  & 0.298\\
IRMOF-12     & 0.213 & 33.6  & 0.420  & 0.516 & OTEDOI  & 0.169 & 22.4  & 0.585  & 0.311\\
IRMOF-14     & 0.206 & 35.4  & 0.376  & 0.514 & QOXVEH  & 0.172 & 22.7  & 0.584  & 0.307\\
IRMOF-15     & 0.209 & 37.1  & 0.354  & 0.841 & QOXSUU  & 0.164 & 22.1  & 0.578  & 0.265\\
IRMOF-20     & 0.230 & 26.0  & 0.655  & 0.456 & RIXPIZ  & 0.173 & 23.3  & 0.570  & 0.305\\
MUZKEZ       & 0.195 & 32.5  & 0.406  & 0.442 & OTEDUO  & 0.170 & 23.1  & 0.565  & 0.297\\
GUFYUE       & 0.178 & 22.6  & 0.609  & 0.452 & WOFQEQ  & 0.177 & 23.9  & 0.565  & 0.297\\
QOXTOP       & 0.151 & 18.5  & 0.669  & 0.234 & QOXRIH  & 0.173 & 23.5  & 0.563  & 0.314\\
             &       &       &        &       & QOXRAZ  & 0.175 & 24.0  & 0.555  & 0.307\\
\hline
\end{tabular}
\end{center}
\end{table}

Among the classical MOF set, IRMOF-20 has the highest volumetric 
capacity, 0.230 kg/L, and IRMOF-15 has the highest gravimetric 
capacity, 37.1 wt. \%. The lowest gravimetric capacities correspond to 
QOXTOP MOF, among selected Al-based MOFs, and to IRMOF-5, among the 
classical MOF collection. All these results are also in agreement with 
the results obtained for the hydrogen case.

\begin{figure}[htb]
\vspace{0.0cm}
\begin{center}
\leavevmode
\put(-220,   0){%
\includegraphics[width=7cm]{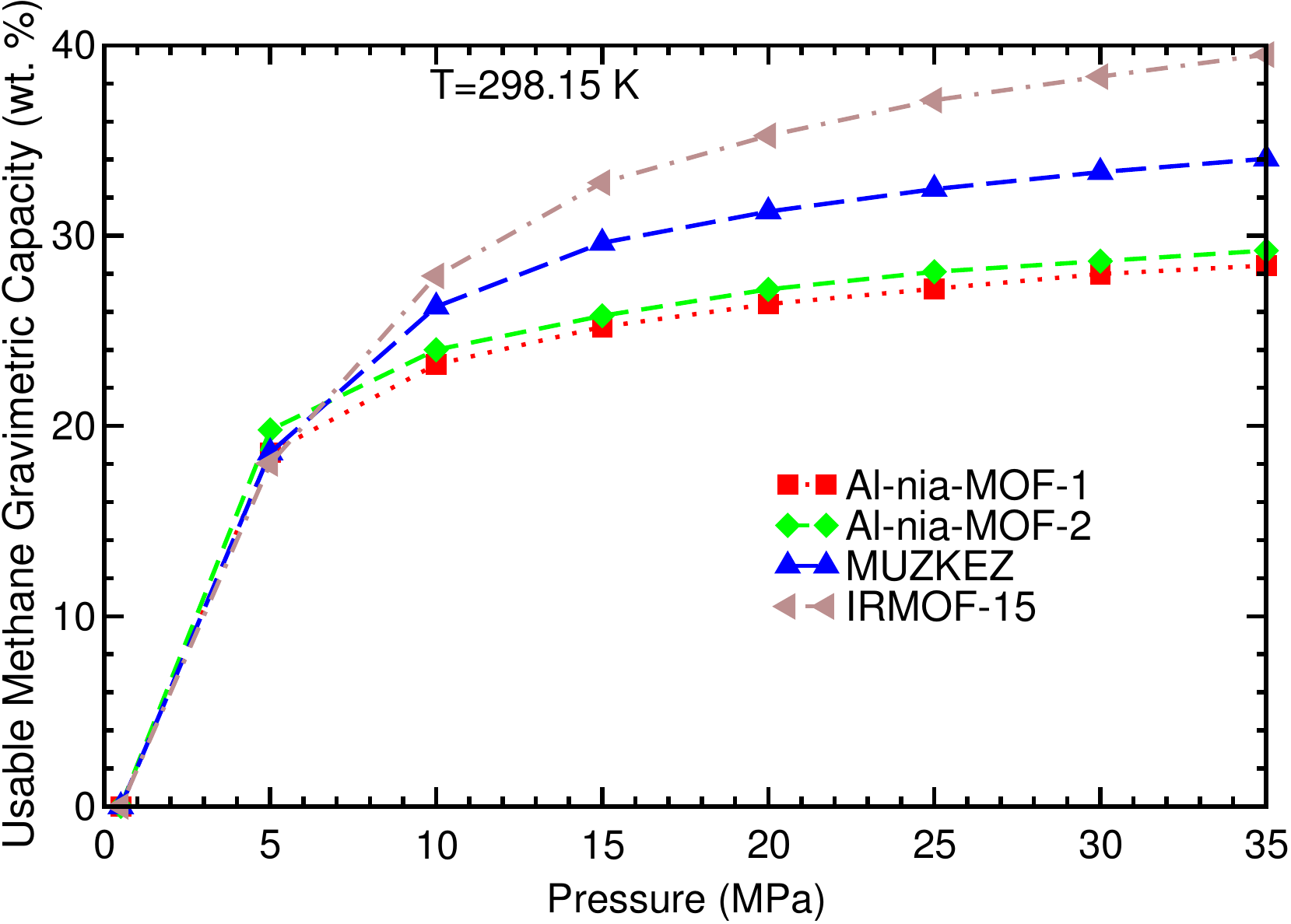}
}
\put(   0,   0){%
\includegraphics[width=7.1cm]{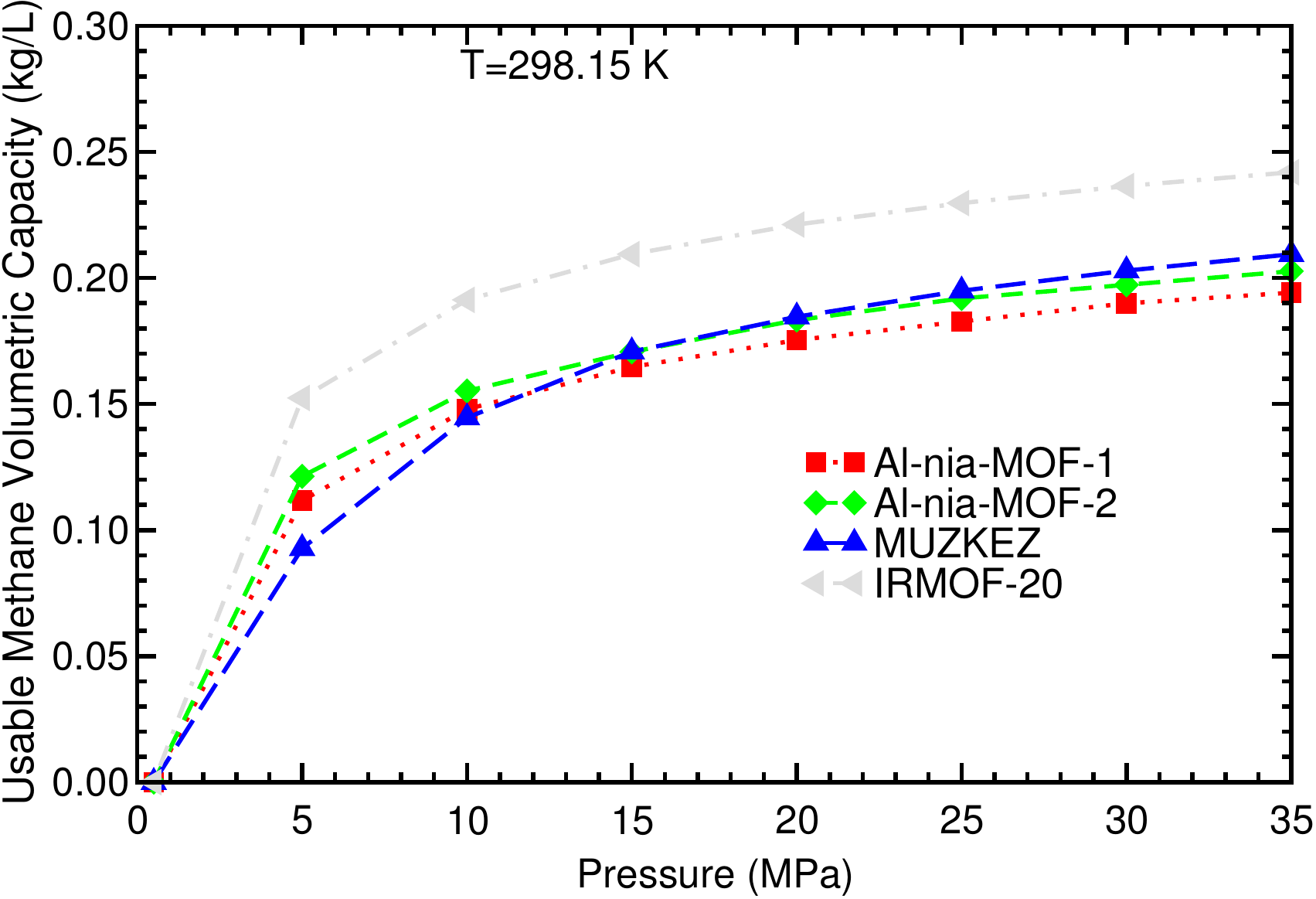}
}
\end{center}
\vspace{-0.5cm}
\caption{Methane volumetric and gravimetric usable capacities vs 
pressure at room temperature of some selected MOFs and the two novel 
Al-nia MOFs.}
\label{ch4cap}
\end{figure}

\begin{figure}[htb]
\vspace{0.0cm}
\begin{center}
\leavevmode
\put(-220,   0){%
\includegraphics[width=7cm]{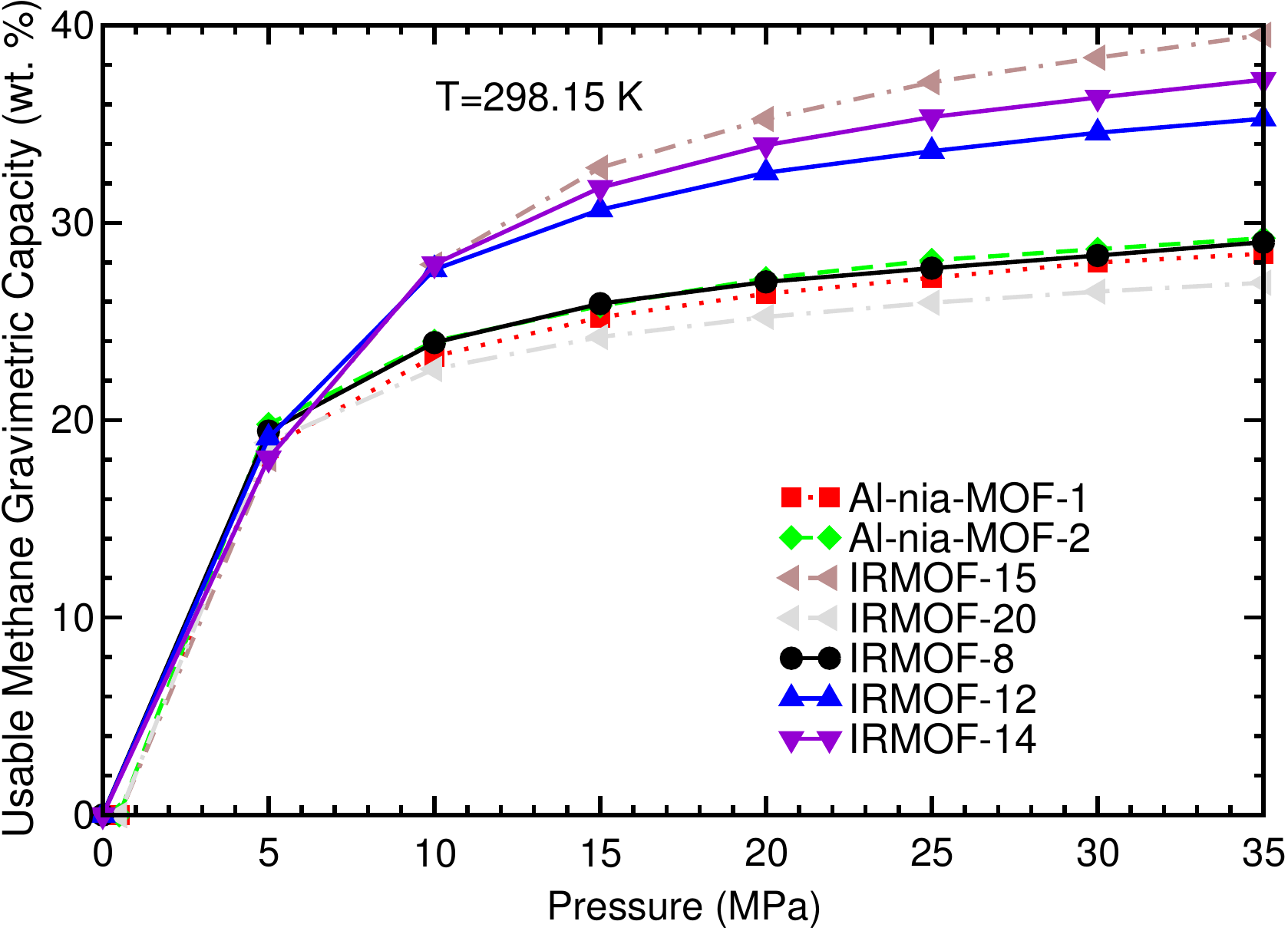}
}
\put(   0,   0){%
\includegraphics[width=7.1cm]{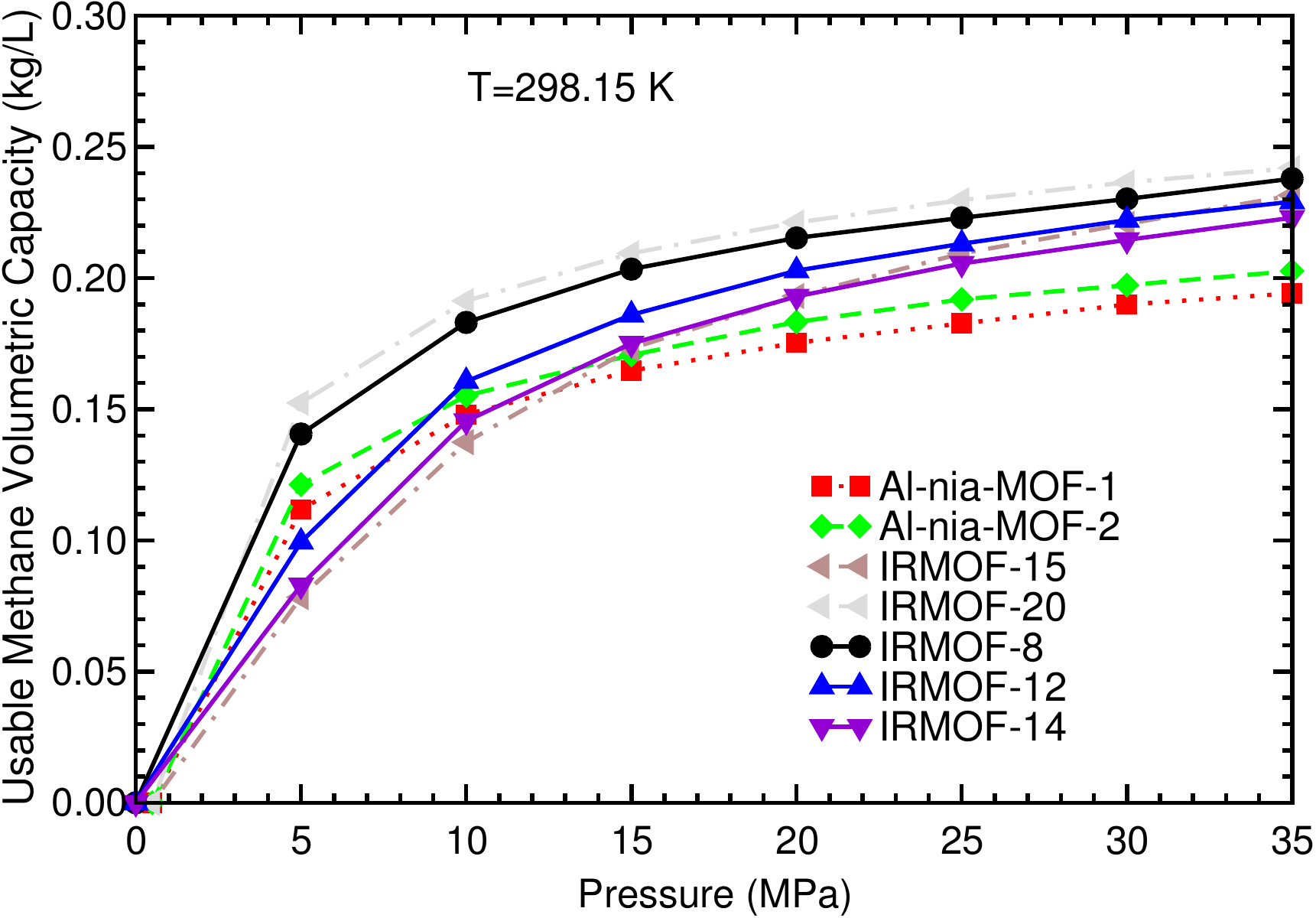}
}
\end{center}
\vspace{-0.5cm}
\caption{
Methane volumetric and gravimetric usable capacities at room temperature vs 
pressure of some selected classical MOFs and the two novel Al-nia-MOFs.}
\label{ch4classics}
\end{figure}

The usable methane gravimetric and volumetric capacities of selected 
Al-based, classical MOFs and Al-nia MOFs are compared on 
Fig.~\ref{ch4cap} and~\ref{ch4classics}, as functions of the pressure in a range between 0.5 
and 35 MPa. Usable gravimetric capacities are quite different. The 
gravimetric storage of all the MOFs increase very fast at low pressures 
and all of them start to saturate approximately at 10 MPa. IRMOF-15 has 
the largest gravimetric capacity from 8 MPa, approximately, and at room 
temperature. Al-nia MOFs, IRMOF-15 and MUZKEZ have similar gravimetric 
capacities in the pressure interval of 0.5-8 MPa, but Al-nia-MOF-2 has 
the highest one. MUZKEZ MOF gravimetric capacity is approximately in 
the middle, between the IRMOF-15 and Al-nia MOFs capacities, which have 
lower values. Al-nia MOFs do not reach the DOE targets for methane, but 
they are very close. IRMOF-15 reaches the gravimetric target when 
pressure gets to 16 MPa and MUZKEZ when it reaches 30 MPa. On the other 
hand, volumetric capacities are very similar. The volumetric capacities 
for Al-nia and MUZKEZ MOFs are slightly smaller than the volumetric 
capacity of IRMOF-20.

The methane storage capacities of the two novel Al-nia MOFs are high at 
25-35 MPa and are close to the capacities of the best classical and 
Al-based MOFs. This is due to their low density and high porosity (See 
Fig.~\ref{ch4cap} and Table~\ref{ch4captab}).

A comparison of the storage capacities of the Al-nia-MOFs and the classical 
MOFs with the highest methane storage capacities at 25 MPa: IRMOF-8, 12, 14, 
15, and 20, was also carried out and is depicted in Fig.~\ref{ch4classics}. 
The gravimetric capacities of the two Al-nia-MOFs, IRMOF-8 and IRMOF-20 are 
relatively similar and smaller than the capacities of the other MOFs. The 
volumetric capacities of the classical MOFs are very similar and higher than 
the volumetric capacities of the Al-nia-MOFs.

\begin{figure}[htb]
\vspace{0.0cm}
\begin{center}
\leavevmode
\put(-220, 150){%
\includegraphics[width=7cm]{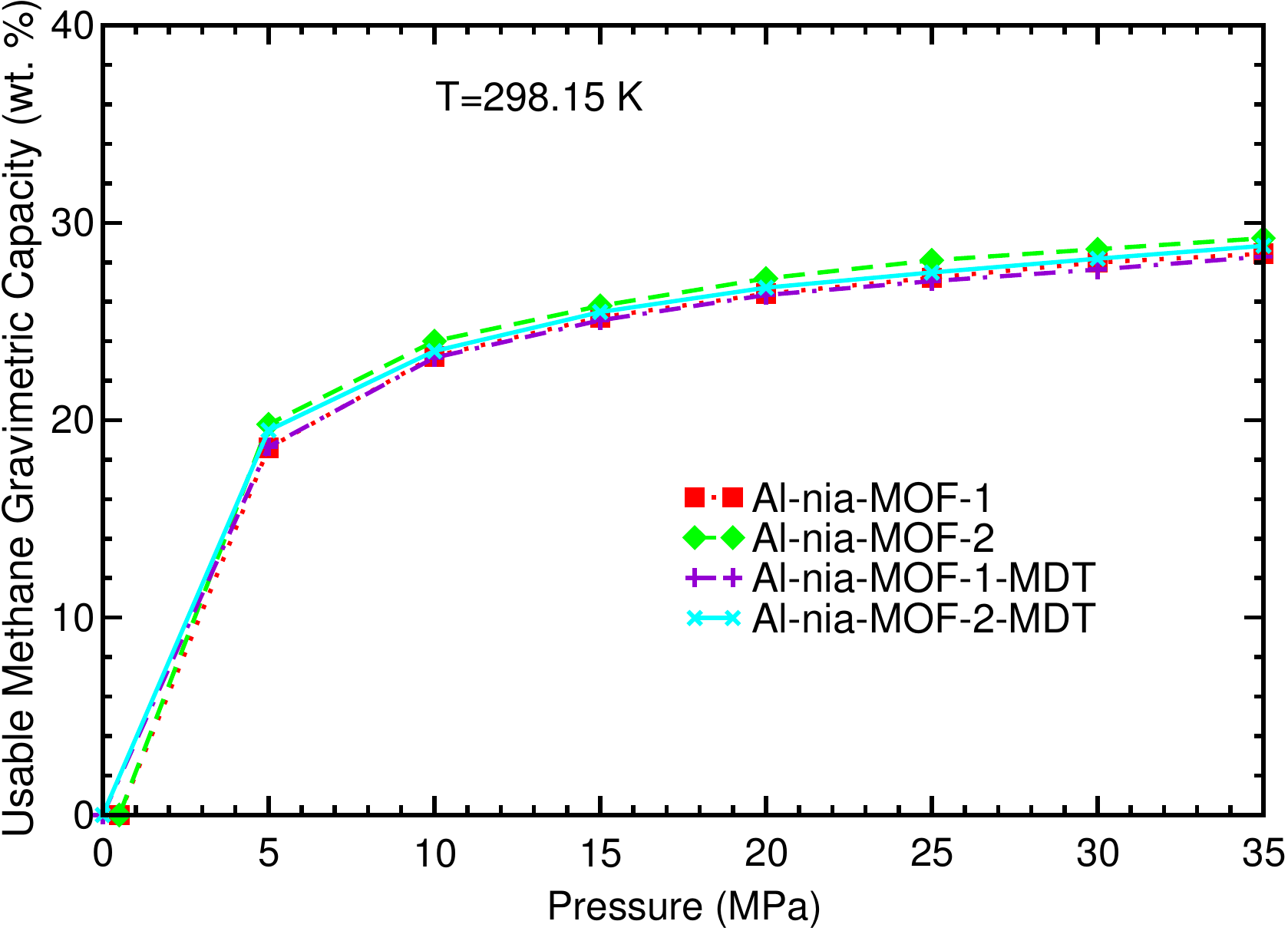}
}
\put(   0, 150){%
\includegraphics[width=7cm]{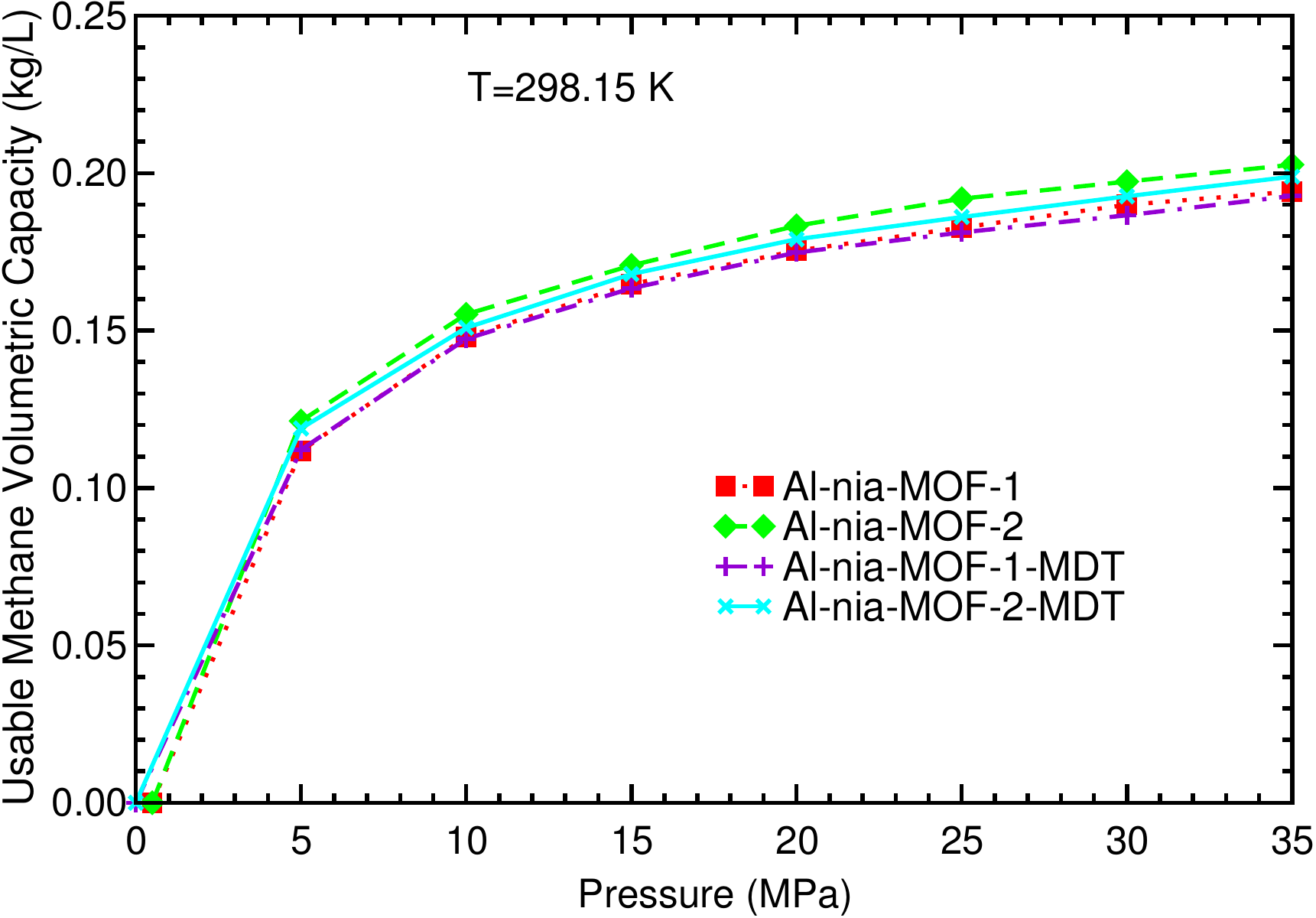}
}
\put(-220,   0){%
\includegraphics[width=7cm]{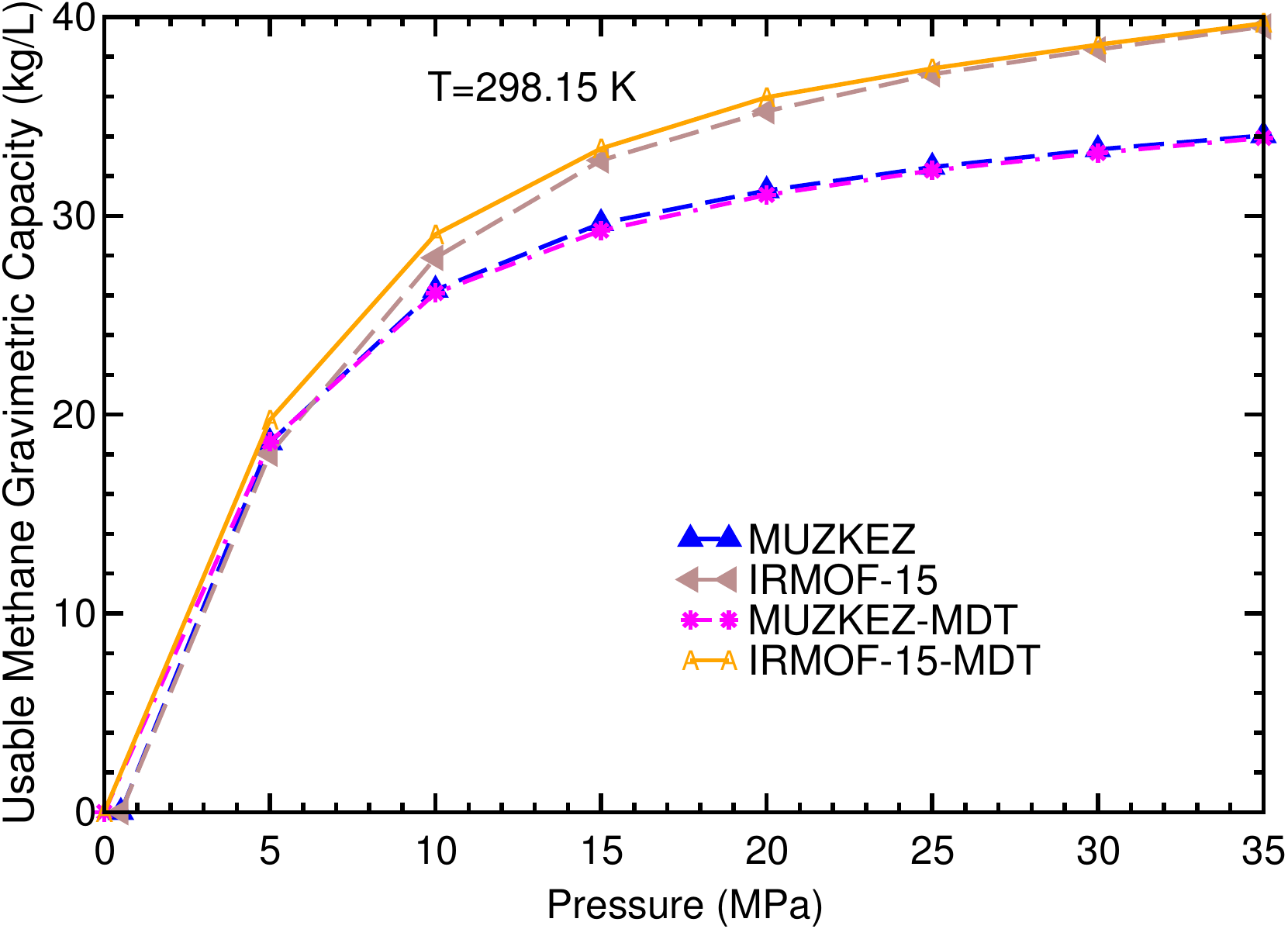}
}
\put(   0,   0){%
\includegraphics[width=7cm]{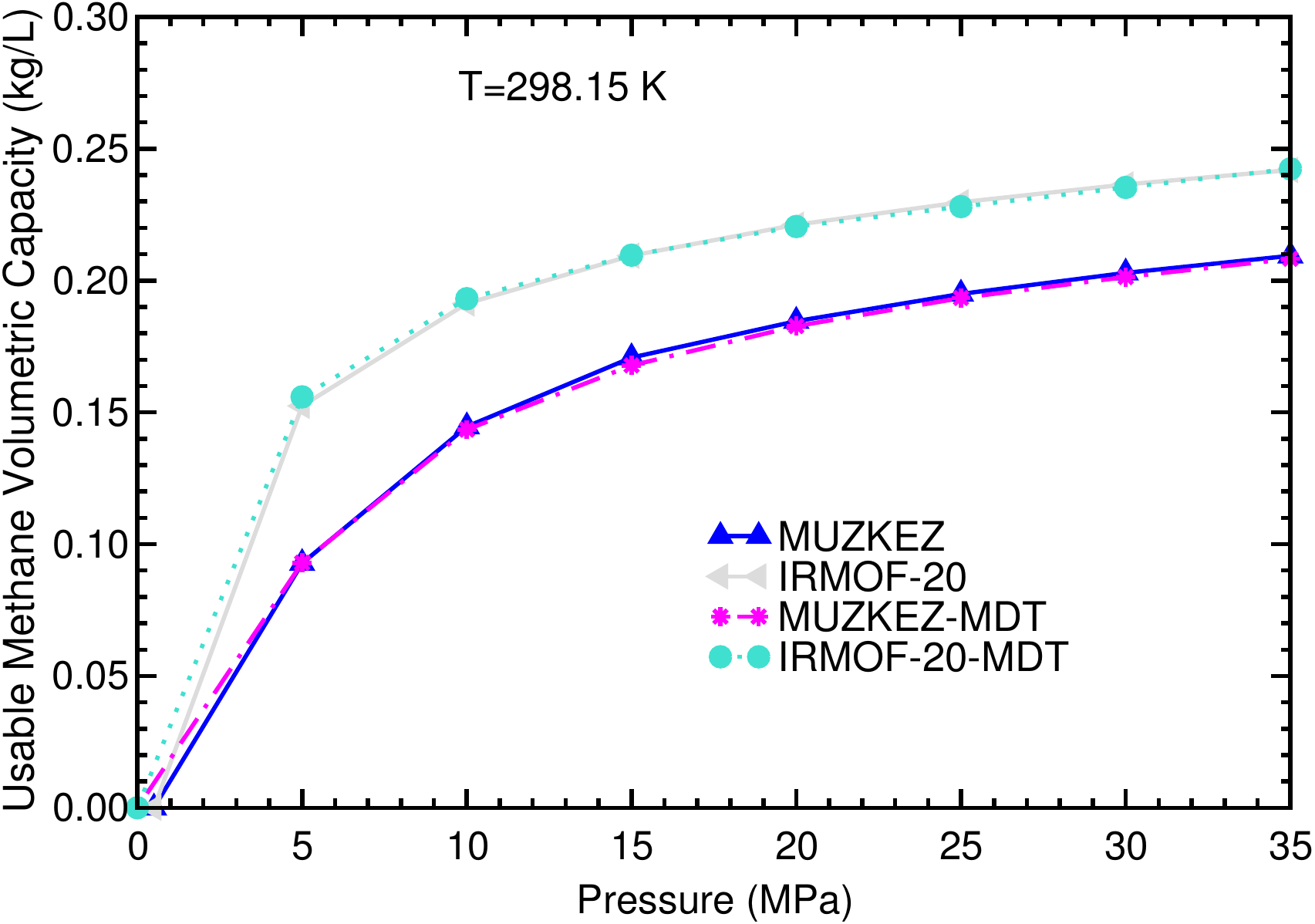}
}
\end{center}
\vspace{-0.5cm}
\caption{
Methane gravimetric and volumetric usable capacity vs pressure at room 
temperature of the Al-nia-MOFs (upper panels) and the best Al-based MOF, 
MUZKEZ, and the best classical MOF for gravimetric, IRMOF-15, and for 
volumetric capacity, IRMOF-20 (lower panels).}
\label{ch4MDTset}
\end{figure}

Similar to the hydrogen case, simulations have been conducted to calculate 
the methane storage capacities with the MDT set of LJ coefficients (See 
Table~\ref{coeffsLJMDT}). The resulting storage capacities at room 
temperature have been depicted in Fig.~\ref{ch4MDTset} as functions of 
pressure. Remarkably, both Al-nia-MOFs and Al-nia-MOFs-MDT showcase almost 
identical capacities. The same applies to the best selected Al-based MOFs 
and classical MOFs studied (MUZKEZ, GUFYUE, IRMOF-15 and IRMOF-20). This 
close similarity is mainly due to the nearly identical LJ coefficients of 
the methane molecule (See Tables~\ref{coeffsLJ} and~\ref{coeffsLJMDT}).

Alezi et al. \cite{alniamof2022} have found that Al-nia-MOF-1 has a 
slightly higher porosity than Al-nia-MOF-2 \cite{alniamof2022}, in 
contrast with the present GCMC simulations: Al-nia-MOF-2 presents a 
slightly higher porosity than Al-nia-MOF-1, 0.416 and 0.340 for 
methane, respectively (See Table~\ref{ch4captab}). Alezi et al. 
\cite{alniamof2022} investigated \ch{CH4} adsorption on Al-nia-MOF-1 at 
8 MPa and different temperatures. The storage capacities obtained by 
Alezi et al.\ and the theoretical capacities obtained in the present 
GCMC simulations are shown and compared in Table~\ref{theovsexp}.

\begin{table}[ht]
\begin{center}
\caption{\noindent
\label{theovsexp}
Experimental and present GCMC total and usable methane storage 
capacities of Al-nia-MOF-1. Pressure is in MPa, temperatures are in K, 
v$_c$ in kg/L and g$_c$ in wt. \% units. The experimental results have 
been published by Alezi et al. \cite{alniamof2022}.}
\bigskip
\begin{tabular}{cccccc}
\hline
P  & T   & v$_c$ & g$_c$ & Capacities & Source\\
\hline
 8 & 258 & 0.188 &  28.6 & Total    & Experiments\\
 8 & 258 & 0.201 &  29.1 & Total    & GCMC\\
 8 & 258 & 0.199 &  28.9 & Total    & GCMC MDT\\
\hline
 8 & 273 & 0.166 &  24.5 & Total    & Experiments\\
 8 & 273 & 0.187 &  27.6 & Total    & GCMC\\
 8 & 273 & 0.182 &  27.2 & Total    & GCMC MDT\\
\hline
 8 & 298 & 0.142 &  23.1 & Total    & Experiments\\
 8 & 298 & 0.161 &  24.8 & Total    & GCMC       \\
 8 & 298 & 0.159 &  24.5 & Total    & GCMC MDT\\ 
\hline
 8 & 298 & 0.123 &  21.3 & Usable   & Experiments\\
 8 & 298 & 0.139 &  22.1 & Usable   & GCMC\\
 8 & 298 & 0.138 &  22.0 & Usable   & GCMC MDT\\ 
\hline
\end{tabular}
\end{center}
\end{table}

The theoretical GCMC and experimental volumetric and gravimetric 
storage capacities are quite similar. The theoretical capacities are 
about 6-14 \% higher than the experimental ones. On the other hand, the 
theoretical capacities from the MDT set are very similar to the theoretical 
capacities obtained using the original set of LJ values and slightly closer 
to the experimental results. The difference between the values of $\sigma$ and 
$\epsilon$ for methane of the two sets of LJ values is very small and this 
causes, in turn, small differences between the methane capacities obtained 
using the two sets of LJ values. The usable capacities reported by Alezi 
et al. \cite{alniamof2022} are estimated, not measured values. This 
comparison with the experiments is very relevant and allow us to estimate 
the relative error of the present GCMC predictions of the storage capacities.

The GCMC simulations are based on the experimental MOF structure and 
hence, on the crystal density, but the density of the samples of MOFs 
used in the experiments is the packing density, which is usually larger 
than the crystal density. The smaller crystal density implies that the 
theoretical storage capacities are larger than the experimental storage 
capacities.


\section{Conclusions and future lines}

The present GCMC results are predictions of the usable hydrogen and 
methane storage capacities of two novel Al-nia MOFs, Al-nia-MOF-1 and 
Al-nia-MOF-2, at room temperature and pressures between 0.5 and 35 MPa, and 
they could be useful for experimentalists. 

According to the present GCMC simulations, the two novel Al-nia MOFs 
show high hydrogen and methane storage capacities at room temperature 
and pressures of 25-35 MPa, comparable to those of the best classical 
and Al-based MOFs. The usable hydrogen storage capacities of these two 
MOFs are about 3 wt. \% and 0.016 kg/L and their usable methane 
storage capacities are about 28 wt. \% and 0.190 kg/L at room 
temperature and 25-35 MPa.

The methane storage capacities of Al-nia-MOF-1 obtained in the GCMC 
simulations at room temperatures (258, 273 and 298 K) and 8 MPa are 
very similar to the experimental values at the same temperatures and 
pressure reported by Alezi et al. \cite{alniamof2022}. The obtained 
GCMC methane storage capacities are about 6-14 \% higher than the 
experimental values.

The dependence of the hydrogen volumetric and gravimetric storage 
capacities of the MOFs studied (classical, Al-based and the two novel 
MOFs) on the porosity and density has been studied. The analysis 
reveals that the capacities are inversely proportional to the density 
and proportional to the porosity. The origin of the high capacities of 
the novel MOFs are their high porosity and low density and these 
capacities could be improved by doping these novel MOFs with some light 
elements.

\medskip
\bigskip
\textbf{Acknowledgment}
\bigskip
This work was founded by MICINN research project from Spain (Grant 
PGC2018-093745-B-I00), Junta de Castilla y Le\'on (Grant VA124G18) and 
the University of Valladolid, Spain. The use of the computer facilities 
of Centro de Proceso de Datos - Parque Cient\'ifico of the University of 
Valladolid is acknowledged.




\end{document}